\documentclass[pra,12pt,tightenlines,superscriptaddress,eqsecnum,nofootinbib,secnumarabic]{revtex4}

\usepackage{bm,amssymb,amsmath,amsthm,mathrsfs,enumerate,verbatim}

\theoremstyle{plain}
\newtheorem{thm}{Theorem}[section]
\newtheorem{lem}[thm]{Lemma}
\newtheorem{cor}[thm]{Corollary}
\newtheorem{prp}[thm]{Proposition}

\theoremstyle{definition}
\newtheorem{dfn}[thm]{Definition}

\theoremstyle{remark}

\DeclareSymbolFont{AMSb}{U}{msb}{m}{n}
\DeclareMathSymbol{\N}{\mathbin}{AMSb}{"4E}
\DeclareMathSymbol{\Z}{\mathbin}{AMSb}{"5A}
\DeclareMathSymbol{\R}{\mathbin}{AMSb}{"52}
\DeclareMathSymbol{\Q}{\mathbin}{AMSb}{"51}
\DeclareMathSymbol{\I}{\mathbin}{AMSb}{"49}
\DeclareMathSymbol{\C}{\mathbin}{AMSb}{"43}
\DeclareMathSymbol{\F}{\mathbin}{AMSb}{"46}
\DeclareMathSymbol{\E}{\mathbin}{AMSb}{"45}

\DeclareSymbolFont{symbolsC}{U}{txsyc}{m}{n}
\DeclareMathSymbol{\coloneq}{\mathrel}{symbolsC}{66}
\DeclareMathSymbol{\eqcolon}{\mathrel}{symbolsC}{67}

\def\rank{\operatorname{rank}}
\def\tr{\operatorname{tr}}
\def\ket#1{|#1\rangle}
\def\bra#1{\langle#1|}
\def\ketbra#1{| #1 \rangle\langle #1 |}
\def\braket#1#2{\langle#1|#2\rangle}
\def\Ket#1{|#1)}
\def\Bra#1{(#1|}
\def\KetBra#1{|#1)(#1|}
\def\BraKet#1#2{(#1|#2)}

\def\End{\operatorname{End}}
\def\Span{\operatorname{span}}
\def\supp{\operatorname{supp}}
\def\ri{\operatorname{ri}}
\def\d{\mathrm{d}}
\def\Qu{\operatorname{Q}}
\def\GC{\mathscr{Q}^\mathrm{gc}}
\def\UC{\mathscr{Q}^\mathrm{uc}}
\def\Is{\mathrm{\bf I}}
\def\Tr{\operatorname{Tr}}
\def\Ho{\operatorname{H}_0}
\def\H{\operatorname{H}}
\def\Hgc{\operatorname{H}^\mathrm{gc}}
\def\Huc{\operatorname{H}^\mathrm{uc}}
\def\Hogc{\operatorname{H}_0^\mathrm{gc}}
\def\Houc{\operatorname{H}_0^\mathrm{uc}}
\def\muu{\mu}
\def\Ptr{\bm{\Pi}_0}
\def\Ps{\bm{\Pi}}
\def\U{\operatorname{U}}
\def\PU{\operatorname{PU}}
\def\PC{\operatorname{PC}}
\def\Wg{\operatorname{Wg}}
\def\Fs{\mathcal{F}}
\def\Fsi{\tilde{\mathcal{F}}}
\def\Qs{\mathscr{Q}}
\def\Hs{\mathcal{H}}

\def\S{\mathrm{S}}

\begin{document}

\title{Optimizing quantum process tomography with unitary $\bm{2}$-designs}

\author{A. J. Scott}
\email{andrew.scott@griffith.edu.au}
\affiliation{Centre for Quantum Computer Technology and Centre for Quantum Dynamics,
Griffith University, Brisbane, Queensland 4111, Australia}

\begin{abstract}
We show that weighted unitary 2-designs define optimal measurements on the system-ancilla 
output state for ancilla-assisted process tomography of unital quantum channels. 
Examples include complete sets of mutually unbiased unitary-operator bases. Each of these 
specifies a minimal series of optimal orthogonal measurements. General quantum channels 
are also considered.
\end{abstract}


\maketitle

\section{Introduction}
\label{sec:intro}

Fundamental to the fabrication of quantum information processing devices~\cite{Nielsen00}, such as quantum teleporters, 
key distributers, cloners, gates, and indeed, quantum computers, is the ability to precisely determine an unknown 
transformation on a quantum system. Quality assurance requires a complete characterization of these devices, which can be 
accomplished through a procedure known as {\em quantum process tomography}~\cite{Paris04}: for judicious choices of initial system states, the 
transformation is uniquely identified by the outcomes of measurements on the transformed states. 

The approach of {\em ancilla-assisted\/} quantum process tomography~\cite{Leung03,DAriano01} 
is to encode all information about the transformation into a single bipartite system-ancilla quantum state, and thus completely 
reduce the problem to that of quantum state tomography~\cite{Paris04}. A sequence of measurements on identically 
prepared copies of this  state will then reveal the particular transformation under examination. It is known that, for linear 
tomographic reconstructions of general quantum states, the most robust measurements against statistical error are 
described by tight informationally complete positive-operator-valued measures (tight IC-POVMs)~\cite{Scott06}. Such measures
derive their name from a related concept in frame theory, called a tight frame~\cite{Daubechies86}, and are equivalent to 
weighted complex projective 2-designs~\cite{Delsarte77,Neumaier81,Hoggar82,Scott06}.  

The nonselective evolution of an open quantum system is described by a completely positive, trace preserving, linear 
transformation on quantum states. Such transformations are called {\em quantum channels\/} within the context of quantum 
information theory~\cite{Nielsen00}, as they also describe the degradation of information encoded in quantum states 
after transmission through a noisy communication channel. A {\em unital\/} quantum channel is one which fixes the maximally 
mixed state. These include all probabilistic applications of unitary operators, and thus, within the context of process tomography, 
form the relevant subclass of channels describing closed-system quantum dynamics.

In this article we study ancilla-assisted process tomography of general and unital quantum channels. The possible 
system-ancilla output states of relevance are then naturally housed in proper convex subsets of the set 
of all quantum states, and thus permit optimizations of the measurement over that necessary to identify a general 
quantum state. We find that the most robust measurements against statistical error, when they exist, are again 
described by tight POVMs, though a generalization thereof. In the unital case, these POVMs are equivalent to 
{\em weighted unitary 2-designs\/}~\cite{Dankert06,Gross06}, but in the general case, 
they are shown not to exist.

The article is organized as follows. Sections~\ref{sec:process} and \ref{sec:icpovms} review 
quantum process and state tomography, respectively, paying particular attention to the pertinent case of 
ancilla-assisted process tomography of quantum channels. Section~\ref{sec:optimal} generalizes results of
Ref.~\cite{Scott06}, characterizing the structure of POVMs that are optimal for linear quantum state tomography 
when a member of a convex subset of all possible quantum states need only be distinguished from other members.
In Sec.~\ref{sec:designs} we introduce the concept of a weighted unitary $t$-design, reviewing known results  
and presenting new ones. Finally, in Sec.~\ref{sec:optimalprocess} we make the connection between weighted unitary 
$2$-designs and the POVMs that optimize ancilla-assisted process tomography of unital quantum channels. The article 
then concludes in Sec.~\ref{sec:conclude} where open problems are discussed. In addition, an appendix sets the 
superoperator notation used throughout this article by reviewing a general class of transformations on quantum 
systems called quantum operations.

\section{Quantum process tomography}
\label{sec:process}

The purpose of this article is to optimize the measurements used for ancilla-assisted process tomography of quantum channels, and in particular, 
unital quantum channels. Quantum channels are nonselective quantum operations. The appendix provides some background 
to this broad class of transformations on quantum systems and introduces important concepts and notations relevant to the current study 
of channels. We will proceed by first describing process tomography for unitary operations. This will lead
naturally into that for channels. 

The dynamical evolution of a closed quantum system $\Hs_\mathrm{s}=\C^d$ is described by a unitary operator $U\in\U(d)$. 
In the absence of any physical description of the system, we expect that the Haar probability measure $\mu$ on $\U(d)$ 
most accurately reflects our state of knowledge of $U$. One method to determine $U$ is then to couple $\Hs_\mathrm{s}$ 
to an auxiliary system $\Hs_\mathrm{a}=\C^{d_\mathrm{a}}$ (called the {\em ancilla}) and allow the combined system $\Hs_\mathrm{s}\otimes\Hs_\mathrm{a}$ to 
evolve from some initially known state, $\rho_\mathrm{i}$ say, to 
\begin{equation}
\rho \;=\; (U\otimes I)\, \rho_\mathrm{i} \,(U^{\dag}\otimes I) \;.
\end{equation} 
A measurement on the combined system will then provide information on $U$. By repeating this procedure many times 
over, perhaps on different input states, $U$ can be determined completely. This method of determining quantum dynamics is called 
{\em quantum process tomography}. Although the ancilla could be removed if the initial state were varied, in this article 
we investigate the opposite extreme by choosing $d_\mathrm{a}=d$ and then $\rho_\mathrm{i}=\ketbra{I}$, fixed, where for any $V\in\U(d)$ we define
\begin{equation}\label{eq:ketU}
\ket{V} \;=\; (V\otimes I)\ket{I} \;\coloneq\;\frac{1}{\sqrt{d}} \sum_k V\ket{k}\otimes\ket{k} \;.
\end{equation} 
The pure state $\ketbra{V}$ is a maximally entangled state of $\Hs_\mathrm{s}\otimes\Hs_\mathrm{a}$, and in fact, all maximally entangled states can 
be written in this form. The output state is $\rho=\ketbra{U}$. Note that $U$ can be found from $\ket{U}$ (and vice versa) 
through the relation $\bra{j}U\ket{k}=\sqrt{d}(\bra{j}\otimes\bra{k})\ket{U}$, a special case of the 
Jamio{\l}kowski isomorphism below.

The determination of an unknown unitary $U$ is thus equivalent to the determination of an unknown maximally entangled 
state $\ket{U}$. The latter can be accomplished through quantum state tomography. It is unrealistic, however, to presume that 
each system evolution in the above tomographic procedure can be performed identically. The class of quantum states under 
examination should thus be broadened to include any classical mixture of maximally entangled states: $\rho =\sum_k r_k\ketbra{U_k}$,
where each $r_k>0$ and $\sum_k r_k=1$. This is the output state of a quantum channel.

The (nonselective) evolution of an open quantum system is described by a {\em quantum channel\/}, i.e., a superoperator 
$\mathcal{E}\in\End(\End(\C^d))$ which is both trace preserving and completely positive (see the appendix). 
The channel is said to be {\em unital\/} if it fixes the maximally mixed state: $\mathcal{E}(I)=I$.
Unital channels include all unitary operations $U \odot U^\dag$, and moreover, all random-unitary channels, i.e., 
those which can be implemented by probabilistic applications of unitary operators (as above): 
$\mathcal{E}=\sum_k r_k U_k\odot{U_k}^\dag$. In dimensions $d\geq 3$, however, there exist unital channels which can not be 
decomposed in this way~\cite{Landau93}. Although random-unitary channels are those channels which are most 
relevant to the study of closed-system dynamics, within this context, we will consider the entire class of unital 
channels together. 

The process tomography of a quantum channel follows that for a unitary operator. The output state corresponding to the input 
$\rho_\mathrm{i}=\ketbra{I}$ completely determines the channel:
\begin{equation}\label{eq:Jam}
\rho \;=\; (\mathcal{E}\otimes\mathcal{I})(\rho_\mathrm{i}) \quad\leftrightarrow\quad \mathcal{E} \;=\; d\sum_{j,k,l,m}\tr\!\big[\big(\ket{m}\bra{j}\otimes\ket{l}\bra{k}\big)\rho\big] \, \ket{j}\bra{k}\odot\ket{l}\bra{m} \;.
\end{equation}
This is the so-called {\em Jamio{\l}kowski isomorphism}~\cite{Jamiolkowski72}. It is important that the basis used in the 
right-hand side (RHS) of Eq.~(\ref{eq:Jam}) is that in the definition of $\ketbra{I}$. Note that
\begin{equation}
\tr_\mathrm{s}(\rho) \;=\; \frac{1}{d}\sum_{j,k}\tr[\mathcal{E}(\ket{j}\bra{k})]\,\ket{j}\bra{k}\;=\; \frac{1}{d}\sum_{j,k}\tr[\ket{j}\bra{k}]\,\ket{j}\bra{k} \;=\; \frac{1}{d}\,I \;,
\end{equation}
since all quantum channels are trace preserving, and additionally,
\begin{equation}
\tr_\mathrm{a}(\rho) \;=\; \frac{1}{d}\sum_{j,k}\mathcal{E}(\ket{j}\bra{k})\,\tr[\ket{j}\bra{k}]\;=\; \frac{1}{d}\,\mathcal{E}(I) \;=\; \frac{1}{d}\,I \;,
\end{equation}
for unital channels. This means that the classes of output states of quantum channels, with fixed input $\rho_\mathrm{i}=\ketbra{I}$, do not 
include all types of quantum states. The same could not be said if $\mathcal{E}$ were trace decreasing, i.e.,
a path of a general quantum operation. Denote by
\begin{equation}
\Qu(\Hs) \;\coloneq\; \{A\in\End(\Hs)\,|\, A\geq 0\,,\,\tr(A)=1\}
\end{equation}
the set of all quantum states for $\Hs$. The two convex subsets of $\Qu(\Hs_\mathrm{s}\otimes\Hs_\mathrm{a})$,
\begin{align} 
\GC &\;\coloneq\; \{\rho\in\Qu(\Hs_\mathrm{s}\otimes\Hs_\mathrm{a}) \,|\, \tr_\mathrm{s}(\rho)=I/d\}\;,\text{ and,} \\
\UC &\;\coloneq\; \{\rho\in\Qu(\Hs_\mathrm{s}\otimes\Hs_\mathrm{a}) \,|\, \tr_\mathrm{s}(\rho)=\tr_\mathrm{a}(\rho)=I/d\}\;, 
\end{align}
then correspond to the outputs of, respectively, general and unital quantum channels.

Let $\{\lambda_k\}_{k=0}^{d^2-1}$ be an orthonormal Hermitian operator basis for $\End(\C^d)$ with the choice  
$\lambda_0=I/\sqrt{d}$. The remaining operators, $\lambda_1,\dots,\lambda_{d^2-1}$, then span the $(d^2-1)$-dimensional 
subspace of traceless operators: $\tr(\lambda_k)=\sqrt{d}\tr(\lambda_0\lambda_k)=0$ for all $k>0$. Every quantum state 
for $\Hs=\C^d\otimes\C^d$ is of course expressible in terms of this basis: 
\begin{equation}\label{eq:hermopbasis}
\rho \;=\; \sum_{j,k} r_{jk}\,\lambda_j\otimes\lambda_k \;. 
\end{equation}
The coefficients must be real, $r_{jk}\in\R$, but besides positivity of the state, the only other constraint is from 
normalization: $r_{00}=1/d$. In contrast, the output state $\rho\in\GC$ has $r_{0k}=0$ for all $k>0$, and, in the unital 
case, $\rho\in\UC$ has $r_{k0}=r_{0k}=0$ for all $k>0$. 
The number of outcomes of a measuring instrument capable of identifying one such output from 
within its class of output states can thus be reduced from $d^4$, the number necessary to identify a general quantum state, 
to $d^2(d^2-1)+1$ for general channels, or $(d^2-1)^2+1$ for unital channels. Ancilla-assisted process tomography of 
quantum channels is thus not equivalent to tomographic reconstructions of general system-ancilla quantum 
states.

We will conclude this section by describing how quantum states are naturally embedded in Euclidean space. This approach will later 
provide insight when we harness the concepts of frame theory. Embedded in the complex vector space $\End(\Hs)$ is 
a real vector space of Hermitian operators:
\begin{equation}
\H(\Hs)\coloneq\{A\in\End(\Hs) \,|\, A^\dag=A\}\;.
\end{equation}
Equipped with the Hilbert-Schmidt inner product inherited from $\End(\Hs)$, $\BraKet{A}{B}\coloneq\tr(A^\dag B)$, 
which induces the Frobenius norm, $\|A\|\coloneq\sqrt{\BraKet{A}{A}}$, the vector space $\H(\Hs)$ forms a real 
Hilbert space: $\H(\C^D)\cong\R^{D^2}$. Within the context of ancilla-assisted process tomography it will be assumed that
$\Hs=\Hs_\mathrm{s}\otimes\Hs_\mathrm{a}=\C^d\otimes\C^d\cong\C^D$, i.e., $D=d^2$. The above coefficients $r_{jk}$ then define 
a canonical choice for the isomorphism to $\R^{D^2}$. Define the two subspaces
\begin{align}
\Hgc &\;\coloneq\; \{A\in\H(\C^d\otimes\C^d) \,|\, \tr_1(A)=\tr(A)I/d\} \;<\;\H(\C^D)\;,\text{ and,} \\ 
\Huc &\;\coloneq\; \{A\in\H(\C^d\otimes\C^d) \,|\, \tr_1(A)=\tr_2(A)=\tr(A)I/d\} \;<\; \Hgc \;<\; \H(\C^D) \;, 
\end{align}
which contain the convex sets $\GC$ and $\UC$, respectively, and have dimensions $d^2(d^2-1)+1$ and $(d^2-1)^2+1$.

In general, $\Qu(\Hs)$ is naturally embedded into the vector subspace of $\H(\Hs)$ consisting of
all traceless Hermitian operators:
\begin{equation}
\Ho(\Hs)\coloneq\{A\in\H(\Hs) \,|\, \tr(A)=0\} \;<\; \H(\Hs)\;.
\end{equation}
Define
\begin{equation}\label{eq:proj0}
\Ptr \;\coloneq\; \Is-\frac{1}{D}\,\KetBra{I}
\end{equation}
which projects onto $\Ho(\Hs)$. This projection defines an isometric 
embedding of $\Qu(\C^D)$ into a $(D^2-1)$-dimensional real Hilbert space, 
$\Qu(\C^D)\hookrightarrow\Ho(\C^D)\cong\R^{D^2-1}$, 
\begin{equation}\label{eq:embedding}
\Ket{\rho_0} \;\coloneq\; \Ptr\Ket{\rho} \;=\; \Ket{\rho-I/D} \;,
\end{equation} 
in which the images of pure states lie on a sphere, $\|\ketbra{\psi}-I/D\|=\sqrt{(D-1)/D}$, and the images of mixed states 
within. In the special case $D=2$ the embedding is bijective into this sphere, realizing the Bloch-sphere 
representation of a qubit, but is otherwise only injective. By `isometric' we mean that distances are preserved: 
$\|\rho_0-\sigma_0\|=\|\rho-\sigma\|$. 

It is important to recognize that both $\GC$ and $\UC$ are embedded into proper vector subspaces of 
$\Ho(\C^D)$ under $\Ptr\,$: 
\begin{align}
\GC  & \;\hookrightarrow\; \Hogc \;\coloneq\; \Ptr\Hgc \;=\; \{A\in\H(\C^d\otimes\C^d) \,|\, \tr_1(A)=0\} \;,\text{ and,} \\ 
\UC  & \;\hookrightarrow\; \Houc \;\coloneq\; \Ptr\Huc \;=\; \{A\in\H(\C^d\otimes\C^d) \,|\, \tr_1(A)=\tr_2(A)=0\} \;. \label{eq:Houc}
\end{align}
The dimensions of these subspaces are $d^2(d^2-1)$ and $(d^2-1)^2$, respectively.
In Sec.~\ref{sec:optimal} we will show that POVMs corresponding to tight frames on these subspaces, if they exist, 
are uniquely optimal for ancilla-assisted process tomography.

\section{Quantum state tomography}
\label{sec:icpovms}

This section serves as an introduction to quantum state tomography and is adapted from 
Ref.~\cite[Sec.~4]{Scott06}. Instead of using the complex vector space $\End(\C^D)\cong\C^{D^2}$ as a 
backdrop, however, we will use the embedded real vector space of Hermitian operators, $\H(\C^D)\cong\R^{D^2}$.  

The outcome statistics of a quantum measurement on a 
system $\Hs=\C^D$ are described by a {\em positive-operator-valued measure (POVM)}~\cite{Busch96}. That is, an 
operator-valued function defined on a $\sigma$-algebra over a set $\mathscr{X}$ of outcomes, 
$F:\mathfrak{B}(\mathscr{X})\rightarrow\H(\Hs)$, which satisfies (1)~$F(\mathscr{E})\geq 0$ for all 
$\mathscr{E}\in\mathfrak{B}(\mathscr{X})$ with equality if $\mathscr{E}=\emptyset$, (2)~$F(\bigcup_{k=1}^\infty\mathscr{E}_k)=\sum_{k=1}^\infty F(\mathscr{E}_k)$ 
for any sequence of disjoint sets $\mathscr{E}_k\in\mathfrak{B}(\mathscr{X})$, and (3) the normalization constraint 
$F(\mathscr{X})=I$. In this article we always take $\mathfrak{B}(\mathscr{X})$ to be the Borel $\sigma$-algebra.
When a quantum measurement has a countable number of outcomes, the indexed set of POVM elements
$\{F(x)\}_{x\in\mathscr{X}}$ completely characterizes $F$, and is thus often referred to as the 
``POVM.'' We will call such POVMs {\em discrete\/}.

We will need to express an arbitrary POVM $F$ in a standard form. To do this, note that each POVM defines a 
natural scalar-valued {\em trace measure\/} $\tau(\mathscr{E})\coloneq\tr[F(\mathscr{E})]$, which inherits the 
normalization $\tau(\mathscr{X})=D$. Since each matrix element of $F$ is a complex-valued measure which is 
absolutely continuous with respect to the nonnegative finite measure $\tau$, the POVM can be expressed as
\begin{equation}\label{eq:POVD}
F(\mathscr{E}) \;=\; \int_\mathscr{E}\d\tau(x)\, P(x)\;,
\end{equation}
where the {\em positive-operator-valued density (POVD)} $P:\mathscr{X}\rightarrow\H(\Hs)$ is uniquely 
defined up to a set of zero $\tau$-measure. The POVD $P$ is of course the Radon-Nikodym derivative of $F$ with 
respect to $\tau$. Note that  $\tr(P)=1$, $\tau$-almost everywhere. If $P$ also has unit rank, $\tau$-almost everywhere, 
then we call $F$ a {\em rank-one\/} POVM. In the special case of a discrete POVM, $P(x)= F(x)/\tau(x)$. 

An informationally complete POVM $F$~\cite{Prugovecki77,Busch91,Scott06} is one with 
the property that each quantum state $\rho$ is uniquely determined by its measurement statistics, 
$p(\mathscr{E})\coloneq\tr\left[F(\mathscr{E})\rho\right]$. A sequence of measurements on 
copies of a system in an unknown state, enabling an estimate of these statistics, will then reveal the state.
This process is called {\em quantum state tomography\/}.

\begin{dfn}
A POVM $F:\mathfrak{B}(\mathscr{X})\rightarrow\H(\Hs)$ is called  {\em informationally complete 
with respect to\/} $\Qs\subseteq\Qu(\Hs)$ if for each pair of distinct quantum states 
$\rho\neq\sigma\in\Qs$ there exists an event $\mathscr{E}\in\mathfrak{B}(\mathscr{X})$ such that 
$\tr\left[F(\mathscr{E})\rho\right]\neq\tr\left[F(\mathscr{E})\sigma\right]$. A POVM which is informationally complete 
with respect to $\Qu(\Hs)$ is called an {\em informationally complete POVM (IC-POVM)\/}.
\end{dfn}
For an arbitrary POVM $F$, define the Hermitian superoperator $\Fs:\H(\Hs)\rightarrow\H(\Hs)$ by
\begin{equation}\label{eq:framesuperoperator}
\Fs\;\coloneq\; \int_\mathscr{X} \d\tau(x)\,\KetBra{P(x)}\;,
\end{equation}
which is positive and bounded under the left-right action.
The latter follows from the fact that $\Tr(\Fs)=\int_\mathscr{X} \d\tau(x)\,\BraKet{P(x)}{P(x)} \leq D$ for any POVM, with equality only for rank-one POVMs.
The image and coimage of a Hermitian superoperator are equal. We call this vector subspace of $\H(\Hs)$ 
the {\em support} of $\Fs$ and denote it by $\supp(\Fs)$. 
Let $\Span(\Qs)$ denote the subspace of $\H(\Hs)$ spanned by members of $\Qs$, and let 
$\ri(\Qs)$ denote the relative interior of $\Qs$, which is the interior of $\Qs$ as a subset 
of its affine hull. Now consider the following.

\begin{prp}\label{prp:icpovm}
Let\/ $F:\mathfrak{B}(\mathscr{X})\rightarrow\H(\Hs)$ be a POVM. Then\/ $F$ is informationally complete w.r.t.\ 
$\Qs\subseteq\Qu(\Hs)$ if\/ $\Qs\subseteq\supp(\Fs)$. Moreover, if\/ 
$\ri(\Qs)\neq\emptyset$, then\/ $F$ is informationally complete w.r.t.\ 
$\Qs$ if and only if\/ $\Qs\subseteq\supp(\Fs)$.
\end{prp} 

\begin{proof}
Let $\Qs\subseteq\supp(\Fs)$. Then for the distinct quantum states $\rho\neq\sigma\in\Qs$ we have
\begin{equation}
\int_\mathscr{X}\d\tau(x)\,|\tr[P(x)(\rho-\sigma)]|^2 \;=\; \Bra{\rho-\sigma}\Fs\Ket{\rho-\sigma} \;>\;0 \;,
\end{equation}
since $\rho-\sigma\in\supp(\Fs)$, being a vector subspace, and $\rho-\sigma\neq 0$. Thus there must exist an event 
$\mathscr{E}\in\mathfrak{B}(\mathscr{X})$ with    
\begin{equation}
\int_\mathscr{E}\d\tau(x)\,\tr[P(x)(\rho-\sigma)] \;\neq\;0\;, 
\end{equation}
or equivalently, $\tr[F(\mathscr{E})\rho]\neq\tr[F(\mathscr{E})\sigma]$. This means $F$ is informationally complete 
w.r.t.\ $\Qs$.

Now let $F$ be informationally complete w.r.t.\ $\Qs$, let $\ri(\Qs)\neq\emptyset$, and suppose 
$\Qs\nsubseteq\supp(\Fs)$. There must then exist an operator $A\in\Span(\Qs)$, $A\neq 0$, such that 
\begin{equation}
\Bra{A}\Fs\Ket{A}\;=\;\int_\mathscr{X}\d\tau(x)\,|\tr[P(x)A]|^2\;=\;0\;,
\end{equation}
which means $\tr(PA)=0$, $\tau$-almost everywhere. This operator is therefore traceless:
\begin{equation}
\tr(A)\;=\;\tr[F(\mathscr{X})A]\;=\;\int_\mathscr{X}\d\tau(x)\,\tr[P(x)A]\;=\;0\;.
\end{equation}
Now for any $\rho\in\ri(\Qs)$, if $\epsilon>0$ is chosen small enough, then $\sigma=\rho+\epsilon A$ is 
also a member of $\Qs$, and moreover, $\sigma\neq\rho$ with
\begin{equation}
\tr[F(\mathscr{E})\sigma]\;=\;\tr[F(\mathscr{E})\rho]+\epsilon\int_\mathscr{E}\d\tau(x)\,\tr[P(x)A] \;=\; \tr[F(\mathscr{E})\rho]
\end{equation}
for all $\mathscr{E}\in\mathfrak{B}(\mathscr{X})$. Thus $F$ could not have been informationally complete w.r.t.\ 
$\Qs$. We must therefore have $\Qs\subseteq\supp(\Fs)$.
\end{proof}

This proposition is a straightforward but important 
observation. If a quantum state need only be distinguished from other members of a given 
convex subset $\Qs\subseteq\Qu(\Hs)$, e.g. $\GC$ or $\UC$, then 
since the relative interior of any convex set is nonempty, Proposition~\ref{prp:icpovm} enables 
us to focus on POVMs for which $\supp(\Fs)\supseteq\Qs$. Equivalently, it enables 
us to focus on POVMs for which $\supp(\Fs)\geq\Span(\Qs)$. 

Note that any POVM $F$ is informationally complete w.r.t.\  
$\Qs=\supp(\Fs)\cap\Qu(\Hs)$. With this choice there is a standard 
procedure for reconstructing a member, $\rho\in\Qs$, in terms of its measurement outcome statistics, 
$p(\mathscr{E})=\tr\left[F(\mathscr{E})\rho\right]$. Let $\Fsi$ be the unique superoperator with 
$\supp(\Fsi)=\supp(\Fs)$ for which
\begin{equation}\label{eq:FFidentity}
\Fsi\Fs\;=\;\Fs\Fsi\;=\;\Ps_\Fs\;,
\end{equation}
where $\Ps_\Fs$ denotes the projector onto $\supp(\Fs)$.
Of course, $\Fsi=\Fs^{-1}$ when $\Qs=\Qu(\Hs)$ (as in Ref.~\cite{Scott06}). Now defining
\begin{equation}\label{eq:reconstructOVD}
\Ket{R} \;\coloneq\; \Fsi\Ket{P}\;,
\end{equation}
the left-right action on $\Ket{\rho}$ of the identity
\begin{equation}\label{eq:prereconstruct}
\int_\mathscr{X}\d\tau(x)\,\Ket{R(x)}\Bra{P(x)} \;=\; \int_\mathscr{X}\d\tau(x)\,\Fsi\KetBra{P(x)} \;=\;\Fsi\Fs\;=\;\Ps_\Fs \;,
\end{equation}
allows state reconstruction in terms of the measurement statistics:
\begin{equation}\label{eq:reconstruct}
\rho \;=\; \int_\mathscr{X}\d\tau(x)\, \tr[P(x)\rho]R(x) \;=\; \int_\mathscr{X}\tr[\d F(x)\rho]R(x) \;=\; \int_\mathscr{X}\d p(x)R(x)\;.
\end{equation}
Although the reconstruction operator-valued density $R$ is generally not positive, it inherits all other properties of $P$, 
i.e. $\int_\mathscr{X}\d\tau(x)R(x)=I$ and $\tr(R)=1$ (see Ref.~\cite{Scott06}).
Finally, it is straightforward to confirm that
\begin{equation}\label{eq:framesuperoperatorinverse}
\Fsi\;=\; \int_\mathscr{X} \d\tau(x)\,\KetBra{R(x)}\;.
\end{equation}

Although we could now proceed directly to an analysis of optimality, we will first briefly show how to 
embed POVMs into Euclidean space, just as was done for quantum states. First note that, for an arbitrary POVM $F$, the subspace 
$\Ho(\Hs)<\H(\Hs)$ is $\Fs$-invariant, and in fact,
\begin{equation}
\Fs\;=\; \Ptr\,\Fs\,\Ptr + \frac{1}{D}\,\KetBra{I} \;,
\end{equation}
since the identity operator is always a left-right eigenvector with unit eigenvalue:
\begin{equation}\label{eq:identityeigenvector}
\Fs\Ket{I} \;=\; \int_\mathscr{X} \d\tau(x)\,\Ket{P(x)}\BraKet{P(x)}{I}\;=\; \int_\mathscr{X} \d\tau(x)\,\Ket{P(x)}\;=\; \int_\mathscr{X}\,\Ket{\d F(x)} \;=\; \Ket{I}\;.
\end{equation}
Although the projector $\KetBra{I}/D$ is fixed by the normalization of $F$, its complement is free:
\begin{equation}\label{eq:framesuperoperator0}
\Fs_0\;\coloneq\;\Ptr\,\Fs\,\Ptr\;=\; \Fs - \frac{1}{D}\,\KetBra{I} \;=\; \int_\mathscr{X} \d\tau(x)\,\KetBra{P_0(x)}\;,
\end{equation}
where we define
\begin{equation}
\Ket{P_0} \;\coloneq\; \Ptr\Ket{P} \;=\; \Ket{P-I/D} \;.
\end{equation}
It is the superoperator $\Fs_0$ which can be adjusted for optimality. The POVM is thus embedded into $\Ho(\Hs)$: 
\begin{equation}
\Ket{F_0(\mathscr{E})} \;\coloneq\; \Ptr\Ket{F(\mathscr{E})} \;=\; \Ket{F(\mathscr{E})-\tau(\mathscr{E})I/D} \;=\; \int_\mathscr{E}\d\tau(x)\, \Ket{P_0(x)} \;.
\end{equation}
This means $\tr(F_0)=0$ and $F_0(\mathscr{X})=0$. Note that our ``embedding'' is anchored to the trace measure, however,
in that we cannot find $F$ from $F_0$ without knowledge of $\tau$. 



\section{Optimal linear quantum state tomography}
\label{sec:optimal}

In this section we decide which POVMs are the most robust against statistical error for linear  
tomographic reconstructions of quantum states. Our premise is that a member of some subset of all possible quantum states 
needs to be distinguished from other members. Although we will focus on the two convex subsets, $\GC$ and $\UC$, 
being those sets relevant to the process tomography of general and unital quantum channels,  
the following analysis will apply to other convex subsets $\Qs\subseteq\Qu(\Hs)$ with similar properties. 
The choice $\Qs=\Qu(\Hs)$ was considered in Ref.~\cite{Scott06} and the following can be considered a generalization.

Let $F$ be a POVM which is informationally complete w.r.t.\ $\Qs\subseteq\Qu(\Hs)$. 
Throughout this section we assume a linear state-reconstruction formula valid for all $\rho\in\Qs$ of the form 
\begin{equation}\label{eq:linearreconstruct}
\Ket{\rho} \;=\; \int_{\mathscr{X}} \d p(x)\,\Ket{Q(x)} \;=\; \int_{\mathscr{X}} \BraKet{\d F(x)}{\rho}\, \Ket{Q(x)} \;,
\end{equation}
where $Q:\mathscr{X}\rightarrow\Span(\Qs)$ is called a {\em reconstruction OVD\/}. By linearity, this 
formula implicitly presupposes that $\supp(\Fs)\geq\Span(\Qs)$.

It is instructive to start with the special case of a discrete POVM, $\{F(x)\}_{x\in\mathscr{X}}$. Suppose that  
$y_1,\dots,y_N$ are the outcomes of measurements on $N$ identical 
copies of the state $\rho\in\Qs$. One estimate for the outcome probabilities is then 
\begin{equation}
\hat{p}(x)\;=\;\hat{p}(x;y_1,\dots,y_N)\;\coloneq\;\frac{1}{N}\sum_{k=1}^N\delta(x,y_k)\;, 
\label{pestimator}\end{equation}
which gives
\begin{equation}
\hat{\rho} \;=\; \hat{\rho}(y_1,\dots,y_N) \;\coloneq\; \sum_{x\in\mathscr{X}} \hat{p}(x;y_1,\dots,y_N)Q(x)\;,
\label{linearreconstructd}\end{equation}
for an estimate of $\rho$. We will call $\hat{\rho}$ a {\em linear tomographic estimate\/} of $\rho$ to distinguish 
it from more sophisticated choices, such as those from maximum likelihood estimation~\cite{Hradil97,Banaszek99} or 
Bayesian mean estimation~\cite{Jones91,Buzek98,Schack01,Tanaka05,BlumeKohout06}. 

The mean squared Hilbert-Schmidt distance provides a measure of the expected error in our estimate:
\begin{align}\label{eq:error}
e^{(F,Q)}(\rho) &\;\coloneq\; \mathop{\textrm{\large E}}_{y_1,\dots,y_N}\Big[\,\|\rho-\hat{\rho}(y_1,\dots,y_N)\|^2\,\Big] \\
&\;=\;  \sum_{x,y\in\mathscr{X}}\,\mathop{\textrm{\large E}}_{y_1,\dots,y_N}\big[\big(p(x)-\hat{p}(x)\big)\big(p(y)-\hat{p}(y)\big)\big]\BraKet{Q(x)}{Q(y)} \\
&\;=\; \frac{1}{N}\bigg(\sum_{x\in\mathscr{X}}p(x)\BraKet{Q(x)}{Q(x)}\,-\,\tr(\rho^2)\bigg) \\
&\;\eqcolon\; \frac{1}{N}\Big(\Delta_p(Q)-\tr(\rho^2)\Big) \;, \label{esterror}
\end{align}
using Eq.~(\ref{eq:linearreconstruct}) and given that
\begin{equation}
\mathop{\textrm{\large E}}_{y_1,\dots,y_N}\big[\big(p(x)-\hat{p}(x)\big)\big(p(y)-\hat{p}(y)\big)\big]\;=\;\frac{1}{N}\Big(p(x)\delta(x,y)-p(x)p(y)\Big)\;,
\end{equation}
which is an elementary calculation.
Equation~(\ref{esterror}) is also a fitting description 
of the error for a POVM with a continuum of measurement outcomes if we define 
\begin{equation}\label{Delta}
\Delta_p(Q) \;\coloneq\; \int_{\mathscr{X}} \d p(x)\,\BraKet{Q(x)}{Q(x)} 
\end{equation}
in general. This is because a countable partition of the outcome set $\mathscr{X}$ allows any
POVM to be approximated by a discrete POVM. Our estimate $\hat{p}$ 
remains a good approximation for the probability measure $p$, except now with $x$ and $y_1,\dots,y_N$ in 
Eq.~(\ref{pestimator}) indicating members of the partition. In the limit of finer approximating partitions we again arrive 
at Eq.~(\ref{esterror}) for the average error, but now with Eq.~(\ref{Delta}) for $\Delta_p(Q)$. 
Since we have no control over the purity of $\rho$, it is this quantity which is 
now of interest. 

The POVM which minimizes $\Delta_p(Q)$, and hence the error, will depend on the quantum state under examination. 
When $\Qs=\Qu(\C^D)$ it is natural to remove this dependence by averaging over all Hilbert-space orientations 
between the system and measurement apparatus. That is, we set $\rho=\rho(\sigma,U)\coloneq U\sigma U^\dag$ where $\sigma\in\Qu(\C^D)$ is fixed, and 
average $\Delta_p(Q)$ over random choices of $U\in\U(D)$. When $\Qs=\GC,\UC\subseteq\Qu(\Hs_\mathrm{s}\otimes\Hs_\mathrm{a})$ the natural 
procedure is to average over all local Hilbert-space orientations $U_\mathrm{s}\in\mathrm{U}(d)$ between the system and measurement 
apparatus, and all local Hilbert-space orientations $U_\mathrm{a}\in\mathrm{U}(d)$ between the system and ancilla. The end result is the 
same, however. Setting $\rho=\rho(\sigma,U_\mathrm{s}\otimes U_\mathrm{a})= (U_\mathrm{s}\otimes U_\mathrm{a})\sigma (U_\mathrm{s}\otimes U_\mathrm{a})^\dag$ we take the average over all 
$U_\mathrm{s},U_\mathrm{a}\in\U(d)\,$:
\begin{align}
\iint_{\mathrm{U}(d)}&\d\muu(U_\mathrm{s})\d\muu(U_\mathrm{a})\,\Delta_{p}(Q) \nonumber\\
&\;=\; \iint_{\mathrm{U}(d)}\d\muu(U_\mathrm{s})\d\muu(U_\mathrm{a})\int_{\mathscr{X}}\tr\!\big[\d F(x)(U_\mathrm{s}\otimes U_\mathrm{a})\sigma (U_\mathrm{s}\otimes U_\mathrm{a})^\dag\big]\BraKet{Q(x)}{Q(x)} \\
&\;=\; \frac{1}{D}\int_{\mathscr{X}}\tr[\d F(x)]\tr(\sigma)\BraKet{Q(x)}{Q(x)} \\
&\;=\; \frac{1}{D}\int_{\mathscr{X}}\d\tau(x)\,\BraKet{Q(x)}{Q(x)} \\
&\;\eqcolon\; \frac{1}{D}\,\Delta_\tau(Q)\;, \label{DeltaR}
\end{align}
where $\muu$ is the unit Haar measure, using Shur's lemma for the integrals. 

It would be presumptuous to take $\Delta_\tau(Q)$ as an error estimate for an arbitrary subset $\Qs\subseteq\Qu(\C^D)$ 
without further information on its structure. Nevertheless, assume that there is a natural set 
of possible ``orientations'' $\mathscr{O}\subseteq\U(D)$ between $\Hs=\C^D$ and the measuring apparatus, 
and a probability measure $\nu$ on $\mathscr{O}$, with the property that for any $\sigma\in\Qs$,
\begin{equation}\label{eq:orientavg}
\int_{\mathscr{O}}\d\nu(U)\,U\sigma U^\dag \;=\; \frac{1}{D}\, I \;.
\end{equation}
Then $\int_{\mathscr{O}}\d\nu(U)\,\Delta_{p}(Q)=\Delta_\tau(Q)/D$ as above. We thus take $\mathscr{O}=\U(d)\otimes\U(d)$ and 
$\nu=\muu\times\muu$ when $\Qs=\GC$ or $\UC$. Another example is  
\begin{equation}
\Qs^\mathrm{cl} \;\coloneq\; \{\rho\in\Qu(\C^D)\,|\,\text{$\rho$ is diagonal in the standard basis}\} \;. 
\end{equation}
The members of $\Qs^\mathrm{cl}$ might be described as ``classical'' states, being convex combinations of basis states: $\rho=\sum_k r_k\ketbra{k}$. 
Under random permutations $U$ of basis elements, Eq.~(\ref{eq:orientavg}) is satisfied for any $\sigma\in\Qs^\mathrm{cl}$. 
In general, we suspect that the above averaging of the error makes sense whenever $\Qs$ is a convex subset of $\Qu(\C^D)$, 
containing the completely mixed state $I/D$, and possessing a symmetry about this state described by Eq.~(\ref{eq:orientavg}). 

We now proceed to an analysis of optimality. The above considerations are summarized as a definition: 
\begin{equation}\label{eq:averror}
e_\mathrm{av}^{(F,Q)}(\sigma) \;\coloneq\; \int_{\mathscr{O}}\d\nu(U)\:e^{(F,Q)}(\rho(\sigma,U)) \;=\; \frac{1}{ND}\Big(\Delta_\tau(Q)-D\tr(\sigma^2)\Big) \;.
\end{equation}
Our goal now is to find the optimal pairs $(F,Q)$ which minimize $e_\mathrm{av}^{(F,Q)}$ for a given fixed $\Qs$. 

There are generally many different choices for the reconstruction OVD $Q:\mathscr{X}\rightarrow\Span(\Qs)$ that satisfy
Eq.~(\ref{eq:linearreconstruct}). Our next task is to show that the canonical choice we encountered in Sec.~\ref{sec:icpovms} is 
uniquely optimal. We thus minimize $\Delta_\tau(Q)$ over all $Q$ while keeping $F$ fixed. Our only constraint is that 
our state-reconstruction formula [Eq.~(\ref{eq:linearreconstruct})] remains valid 
for all $\rho\in\Qs$. By linearity, this formula is then also valid for any member of $\Span(\Qs)$, and therefore, 
\begin{equation}\label{eq:prereconstructidentity}
\int_{\mathscr{X}} \Ket{Q(x)}\Bra{\d F(x)}\Ps_\Qs \;=\; \int_{\mathscr{X}}\d\tau(x)\, \Ket{Q(x)}\Bra{P'(x)} \;=\; \Ps_\Qs\;,
\end{equation}
where $\Ps_\Qs$ projects onto $\Span(\Qs)$ and $\Ket{P'}\coloneq\Ps_\Qs\Ket{P}$.
This means $\{Q(x)\}_{x\in\mathscr{X}}$ is a dual frame to the frame $\{P'(x)\}_{x\in\mathscr{X}}$, w.r.t.\ $\tau$, 
within the subspace $\Span(\Qs)$. Consult Christensen~\cite{Christensen03} for an introduction to frame theory (see also Ref.~\cite{Scott06}). 
The following lemma shows that there is a unique canonical dual frame which is optimal (see Ref.~\cite[Lemma~16]{Scott06} for a proof).

\begin{lem}\label{lem:lemest1}
Let\/ $\{A(x)\}_{x\in\mathscr{X}}$ be an operator frame w.r.t.\ the measure\/ $\alpha$. 
Then for all dual frames\/ $\{B(x)\}_{x\in\mathscr{X}}$,
\begin{equation}
 \int_{\mathscr{X}}\d\alpha(x)\,\BraKet{B(x)}{B(x)} \;\geq\; \int_{\mathscr{X}}\d\alpha(x)\,\BraKet{\tilde{A}(x)}{\tilde{A}(x)} \;,
\end{equation}
with equality only if\/ $B=\tilde{A}$,\/ $\alpha$-almost everywhere, where\/ $\{\tilde{A}(x)\}_{x\in\mathscr{X}}$ is the 
canonical dual frame, i.e.,\/ $\Ket{\tilde{A}}\coloneq\mathcal{A}^{-1}\Ket{A}$ with\/ $\mathcal{A}\coloneq\int_\mathscr{X}\d\alpha(x)\KetBra{A(x)}$. 
\end{lem}

Let $\Fsi'$ be the inverse of $\Fs'\coloneq\Ps_\Qs\Fs\Ps_\Qs$ in the subspace $\Span(\Qs)$ and define 
$\Ket{R'}\coloneq\Fsi'\Ket{P'}$. This means
\begin{equation}
\Fsi'\;=\; \int_\mathscr{X} \d\tau(x)\,\KetBra{R'(x)}\;.
\end{equation}
Lemma~\ref{lem:lemest1} shows that 
\begin{equation}
\Delta_\tau(Q) \;\geq\; \Delta_\tau(R') \;=\; \Tr(\Fsi') \;,
\end{equation}
with equality only if $Q=R'$, $\tau$-almost everywhere. We thus make this choice and now minimize
the quantity $\Tr(\Fsi')$ over all POVMs. Define $\delta'\coloneq\dim\Span(\Qs)$.

\begin{lem}\label{lemest2}
Let\/ $F:\mathfrak{B}(\mathscr{X})\rightarrow\H(\C^D)$ be a POVM which is informationally complete w.r.t.\  
$\Qs\subseteq\Qu(\C^D)$, where\/ $\ri(\Qs)\neq\emptyset$ and\/ $I\in\Span(\Qs)$. Then
\begin{equation}
\Tr(\Fsi') \;\geq\; \frac{(\delta'-1)^2}{D-1}\,+\,1\;,
\end{equation}
with equality if and only if
\begin{equation}\label{eq:lemest2}
\Fs \;=\; \frac{D-1}{\delta'-1}\,\Ps_{\Qs}\,+\,\frac{\delta'-D}{(\delta'-1)D}\,\KetBra{I} \;.
\end{equation}
\end{lem}
\begin{proof}
Since $F$ is informationally complete w.r.t.\ $\Qs$ and $\ri(\Qs)\neq\emptyset$, by Proposition~\ref{prp:icpovm},
we must have $\supp(\Fs)\geq\Span(\Qs)$. Thus $\Fs'=\Ps_\Qs\Fs\Ps_\Qs$ has $\delta'$ nonzero left-right eigenvalues:
$\lambda_1,\dots,\lambda_{\delta'}>0$. One eigenvalue is fixed at unity, however, since 
$\Fs'\Ket{I}=\Ps_\Qs\Fs\Ps_\Qs\Ket{I}=\Ket{I}$ when $\Ket{I}\in\Span(\Qs)$, given that 
$\Ket{I}$ is always an eigenvector of $\Fs$ [Eq.~(\ref{eq:identityeigenvector})]. 
We thus take $\lambda_1=1$. The remaining eigenvalues satisfy
\begin{equation}
\sum_{k=2}^{\delta'} \lambda_k \;=\; \Tr(\Fs')-1  \;\leq\; \Tr(\Fs)-1  \;\leq \; D-1 \;,
\end{equation} 
given that $\Tr(\Fs')=\Tr(\Ps_\Qs\Fs)\leq\Tr(\Fs)$, with equality if and only if $\supp(\Fs)=\Span(\Qs)$, and
$\Tr(\Fs)\leq D$, with equality if and only if $F$ is a rank-one POVM.
 Under this constraint, it is straightforward to show that 
\begin{equation}\label{eq:lemest2proof1}
\Tr(\Fsi') \;=\; \sum_{k=2}^{\delta'} \frac{1}{\lambda_k}\,+\,1 
\end{equation}
takes its minimum value if and only if $\lambda_2=\dots=\lambda_{\delta'}=(D-1)/(\delta'-1)$, 
or equivalently,
\begin{equation}\label{eq:lemest2proof2}
\Fs' \;=\;\frac{D-1}{\delta'-1}\bigg(\Ps_{\Qs}-\frac{1}{D}\,\KetBra{I}\bigg) \,+\,\frac{1}{D}\,\KetBra{I} \;.
\end{equation}
Note that $\Tr(\Fs')=D$ with this choice, however, which requires $\supp(\Fs)=\Span(\Qs)$. We must therefore have 
$\Fs'=\Fs$ and Eq.~(\ref{eq:lemest2proof2}) is equivalent to Eq.~(\ref{eq:lemest2}). Finally, the minimum value 
of Eq.~(\ref{eq:lemest2proof1}) is $\Tr(\Fsi')= (\delta'-1)\cdot\big((\delta'-1)/(D-1)\big)+1=(\delta'-1)^2/(D-1)+1$.
\end{proof}
 
It is important to recognize that our condition for optimality [Eq.~(\ref{eq:lemest2})] sets $\supp(\Fs)=\Span(\Qs)$.
We now proceed under this assumption, effectively replacing each primed symbol 
above by its unprimed counterpart. Furthermore, optimality requires a 
rank-one POVM. This is because Eq.~(\ref{eq:lemest2}) gives $\Tr(\Fs) = D$, which is
possible only for rank-one POVMs. When $\Qs=\Qu(\C^D)$ we recover the 
tight rank-one IC-POVMs described in Ref.~\cite{Scott06}. Let us use the same terminology here.
\begin{dfn}\label{tightdfn}
Let $F:\mathfrak{B}(\mathscr{X})\rightarrow\H(\C^D)$ be a POVM. Then $F$ is called {\em tight\/} if the 
OVD $\{P_0(x)\}_{x\in\mathscr{X}}$ forms a tight operator frame w.r.t.\ $\tau$ in $\supp(\Fs_0)$, i.e.,
\begin{equation}\label{tightdfneq}
\Fs_0\;\coloneq\;\int_\mathscr{X} \d\tau(x)\,\KetBra{P_0(x)}\;=\;a\Ps_{\Fs_0}\;,
\end{equation}
for some constant $a>0$, or equivalently,  
\begin{equation}\label{tightdfneq2}
\Fs \;=\; a\Ps_{\Fs}+\frac{1-a}{D}\,\KetBra{I}\;.
\end{equation}
\end{dfn}
The constant satisfies $a\leq(D-1)/(\delta-1)$, where $\delta\coloneq\rank(\Fs)$ (left-right rank), 
with equality only for rank-one POVMs. Returning to 
Eq.~(\ref{eq:lemest2}) we see that tight rank-one POVMs are precisely those which are optimal for linear 
quantum state tomography. That is,
\begin{equation}\label{tightdfneq3}
\Fs \;=\; \frac{D-1}{\delta-1}\,\Ps_{\Fs}\,+\,\frac{\delta-D}{(\delta-1)D}\,\KetBra{I} 
\end{equation}
if and only if $F$ is a tight rank-one POVM.
The optimal state-reconstruction formula is given by Eq.~(\ref{eq:reconstruct}), which now takes the form 
\begin{equation}\label{eq:reconstructtight}
\rho \;=\; \frac{\delta-1}{D-1}\int_\mathscr{X}\d p(x)P(x)\;-\; \frac{\delta-D}{D(D-1)}\,I\;,
\end{equation}
since a straightforward calculation of $\Fsi$ from Eq.~(\ref{tightdfneq3}) shows that
\begin{equation}\label{eq:reconstructOVDtight}
R \;=\; \frac{\delta-1}{D-1}\,P\;-\; \frac{\delta-D}{D(D-1)}\,I \;.
\end{equation}
We now restate our findings in a theorem.

\begin{thm}\label{thmest}
Let\/ $F:\mathfrak{B}(\mathscr{X})\rightarrow\H(\C^D)$ be a POVM which is 
informationally complete w.r.t.\ $\Qs\subseteq\Qu(\C^D)$, assumed convex and containing $I/D$, and 
let\/ $\delta'=\dim\Span(\Qs)$. Then for any fixed quantum state\/ $\sigma\in\Qs$,
\begin{equation}
e_\mathrm{av}^{(F,Q)}(\sigma) \;\geq\; \frac{1}{ND}\bigg(\frac{(\delta'-1)^2}{D-1}\,+\,1 -D\tr(\sigma^2)\bigg)\;, 
\end{equation}
for all reconstruction OVDs\/ $Q$. Furthermore, equality occurs if and 
only if\/ $Q=R$ and $F$ is a tight rank-one POVM with\/ $\supp(\Fs)=\Span(\Qs)$. 
\end{thm}

Now consider the worst-case error. The average provides a lower bound:
\begin{align}\label{estwcineq}
e_\mathrm{wc}^{(F,Q)}(\sigma) &\;\coloneq\; \sup_{U\in\mathscr{O}}\:e^{(F,Q)}(\rho(\sigma,U))  \\
  &\;\geq\; e_\mathrm{av}^{(F,Q)}(\sigma)  \\
  &\;\geq\; \frac{1}{ND}\bigg(\frac{(\delta'-1)^2}{D-1}+1-D\tr(\sigma^2)\bigg)\;.
\end{align}
Returning to Eq.~(\ref{esterror}), however, but now with $Q=R$ and $\rho=\rho(\sigma,U)=U\sigma U^\dag$, we find  
\begin{align}
e^{(F,R)}(\rho(\sigma,U)) &\;=\; \frac{1}{N}\bigg(\int_{\mathscr{X}}\d p(x)\,\BraKet{R(x)}{R(x)}\,-\,\tr(\sigma^2)\bigg) \\
&\;=\; \frac{1}{N}\bigg(\frac{1}{D}\bigg(\frac{(\delta-1)^2}{D-1}+1\bigg)\int_{\mathscr{X}}\d p(x)\,-\,\tr(\sigma^2)\bigg) \\
&\;=\; \frac{1}{ND}\bigg(\frac{(\delta-1)^2}{D-1}+1-D\tr(\sigma^2)\bigg)\;,
\end{align}
when $R$ satisfies Eq.~(\ref{eq:reconstructOVDtight}) and $P$ is rank one, 
regardless of orientation $U\in\mathscr{O}\subseteq\mathrm{U}(D)$. Thus given $\delta=\delta'$ [$\supp(\Fs)=\Span(\Qs)$]
when $e_\mathrm{av}^{(F,Q)}$ is minimized, the following is a consequence.

\begin{cor}\label{corest}
Let\/ $F:\mathfrak{B}(\mathscr{X})\rightarrow\H(\C^D)$ be a POVM which is 
informationally complete w.r.t.\ $\Qs\subseteq\Qu(\C^D)$, assumed convex and containing $I/D$, and 
let\/ $\delta'=\dim\Span(\Qs)$. Then for any fixed quantum state\/ $\sigma\in\Qs$,
\begin{equation}
e_\mathrm{wc}^{(F,Q)}(\sigma) \;\geq\; \frac{1}{ND}\bigg(\frac{(\delta'-1)^2}{D-1}\,+\,1 -D\tr(\sigma^2)\bigg)\;, 
\end{equation}
for all reconstruction OVDs\/ $Q$. Furthermore, equality occurs if and 
only if\/ $Q=R$ and $F$ is a tight rank-one POVM with\/ $\supp(\Fs)=\Span(\Qs)$. 
\end{cor}

Tight rank-one POVMs are thus optimal for linear quantum state tomography in both an average and worst-case sense.
In fact, they form the unique class of POVMs that achieve
\begin{equation}\label{wceqav}
e_\mathrm{wc}^{(F,R)}(\sigma) \;=\; e_\mathrm{av}^{(F,R)}(\sigma) \;=\; e^{(F,R)}(\rho(\sigma,U))  \;=\;  \frac{1}{ND}\bigg(\frac{(\delta'-1)^2}{D-1}+1-D\tr(\sigma^2)\bigg)\;.
\end{equation}
The exact structure of these POVMs for $\Qs=\GC$ and $\Qs=\UC$, when they exist, will be explored in detail in 
Sec.~\ref{sec:optimalprocess}. To do so, however, we first need to explain the concept of a ``unitary $t$-design.'' This 
is done in the following section. When $\Qs=\Qu(\C^D)$ we recover the results of 
Ref.~\cite[Theorem~18 and Corollary~19]{Scott06}. Lastly, consider $\Qs=\Qs^\mathrm{cl}$. 
Only $\delta'=D$ dimensions are then spanned, giving $\big(1-\tr(\sigma^2)\big)/N$ for the minimum error. 
In particular, for pure states this is zero.

\section{Unitary $\bm{t}$-designs}
\label{sec:designs}

The extension of spherical $t$-designs~\cite{Delsarte77} to the unitary group was recently considered by 
Dankert~{\it et al}.~\cite{Dankert06} and Gross~{\it et al}.~\cite{Gross06}, and the following definition is 
equivalent to theirs. By a ``unitary $t$-design,'' however, we really mean a {\em projective} unitary 
$t$-design, in that each $e^{i\phi}U\in\U(d)$ should always be identified with $U$. With this in mind, let $U(x)\in\U(d)$ denote a 
representative from the equivalence class of unitaries $x\in\PU(d)=\U(d)/\U(1)$ and let $\muu$ denote the 
Haar measure on $\PU(d)$ with the normalization $\muu(\PU(d))=1$. A countable set $\mathscr{S}$ endowed with a weight function 
$w:\mathscr{S}\rightarrow(0,1]$, where $\sum_{x\in\mathscr{S}}w(x)=1$, will be called 
a {\em weighted set\/} and denoted by the pair $(\mathscr{S},w)$. 

\begin{dfn}\label{dfn:wdesign}
A finite weighted set $(\mathscr{D},w)$, $\mathscr{D}\subset\PU(d)$, is called a {\em weighted $t$-design (in dimension $d$)\/} if 
\begin{equation}\label{eq:wdesign}
\sum_{x\in\mathscr{D}} w(x)\, U(x)^{\otimes t}\otimes \big(U(x)^{\otimes t}\big)^\dag \;=\; \int_{\PU(d)}\!\!\d\muu(x)\,U(x)^{\otimes t}\otimes \big(U(x)^{\otimes t}\big)^\dag \;.
\end{equation}
\end{dfn}

When $w(x)=1/|\mathscr{D}|$ we recover the more common notion of an ``unweighted'' $t$-design. Define 
$T\coloneq\sum_{j,k}\ket{j}\bra{k}\otimes\ket{k}\bra{j}$, which satisfies $\tr[(A\otimes B) T]=\tr(AB)$ and
is called the {\em swap\/} (or {\em transposition\/}) since 
$T\ket{\psi}\otimes\ket{\phi}=\ket{\phi}\otimes\ket{\psi}$. By multiplying Eq.~(\ref{eq:wdesign}) on the right by 
$ I^{\otimes (t-1)}\otimes T\otimes I^{\otimes (t-1)}$ and tracing out the inner pair of subsystems, we can immediately 
deduce that every weighted $t$-design is also a weighted $(t-1)$-design. Repeating this process $t$ times shows that 
the normalization of $w$ is in fact already implied by Eq.~(\ref{eq:wdesign}). 
Unweighted $t$-designs in $\PU(d)$ exist for every $t$ and $d\,$:

\begin{thm}[Seymour and Zaslavsky~\cite{Seymour84}]\label{thm:SZ}
Let\/ $\Omega$ be a path-connected topological space endowed with a measure\/ $\omega$ that is finite and positive with 
full support, and, let\/ $f: \Omega \rightarrow \R^m$ be a continuous, integrable function. Then there 
exists a finite set\/ $\mathscr{X}\subseteq\Omega$ such that 
\begin{equation}\label{eq:SZ}
\frac{1}{|\mathscr{X}|}\sum_{x\in\mathscr{X}} f(x) \;=\; \frac{1}{\omega(\Omega)}\int_\Omega\d\omega(x)\, f(x) \;.
\end{equation}
The size of\/ $\mathscr{X}$ may be any number, with a finite number of exceptions.
\end{thm}
\begin{cor}\label{cor:exist}
For each pair of positive integers\/ $t$ and\/ $d$, and for all sufficiently large\/ $n$, there exist (unweighted) 
unitary\/ $t$-designs in dimension\/ $d$ of size\/ $n$.
\end{cor}
\begin{proof}
Simply let $\Omega=\PU(d)$, $\omega=\muu$, and apply Theorem~\ref{thm:SZ} to 
\begin{equation}
f(x) \;\coloneq\; U(x)^{\otimes t}\otimes \big(U(x)^{\otimes t}\big)^\dag\;,
\end{equation}
which maps $\PU(d)$ into $\End(\C^d)^{\otimes 2t}\cong\R^{2d^{4t}}$.
\end{proof}

The task of finding $t$-designs is facilitated by the following theorem. Define the positive constant
\begin{equation}\label{eq:gamma}
\gamma(t,d) \;\coloneq\; \int_{\PU(d)}\!\!\d\muu(x)\, |\tr[U(x)]|^{2t}\;.
\end{equation}

\begin{thm}\label{thm:welch}
For any finite weighted set\/ $(\mathscr{S},w)$,\/ $\mathscr{S}\subset\PU(d)$, and any\/ $t\geq 1$,
\begin{equation}\label{eq:welchbound}
\sum_{x,y\in\mathscr{S}}w(x)w(y)\,|\tr[U(x)^\dag U(y)]|^{2t}\;\geq\; \gamma(t,d) \;,
\end{equation}
with equality if and only if\/ $(\mathscr{S},w)$ is a weighted\/ $t$-design.
\end{thm}
\begin{proof}
Defining $S\coloneq\sum_{x\in\mathscr{S}}w(x)\,U(x)^{\otimes t}\otimes (U(x)^{\otimes t})^\dag-\int_{\PU(d)}\d\muu(x)\,U(x)^{\otimes t}\otimes (U(x)^{\otimes t})^\dag$ 
we see that
\begin{align}
0 \:\leq\: \tr(S^\dag S)& \:=\: \sum_{x,y\in\mathscr{S}}w(x)w(y)\,|\tr[U(x)^\dag U(y)]|^{2t}- 2\sum_{x\in\mathscr{S}}w(x) \int_{\PU(d)}\!\!\d\muu(y)\, |\tr[U(x)^\dag U(y)]|^{2t} \nonumber\\ 
& \qquad\qquad\qquad + \int_{\PU(d)}\!\!\d\muu(x)\int_{\PU(d)}\!\!\d\muu(y)\,|\tr[U(x)^\dag U(y)]|^{2t} \\
& \:=\: \sum_{x,y\in\mathscr{S}}w(x)w(y)\,|\tr[U(x)^\dag U(y)]|^{2t} - \int_{\PU(d)}\!\!\d\muu(x)\, |\tr[U(x)]|^{2t} 
\end{align}
with equality if and only if $S=0$, which is the defining property of a $t$-design.
\end{proof}

This theorem allows us to check whether a weighted subset of $\PU(d)$ forms a $t$-design by considering only the 
``angles'' between the supposed design elements. It also shows that $t$-designs can be found numerically by 
parametrizing a weighted set and minimizing the LHS of Eq.~(\ref{eq:welchbound}). The lower bound can be considered 
a variation on the Welch bound~\cite{Welch74}. The constant $\gamma$ was calculated by Diaconis and Shahshahani~\cite{Diaconis94} 
for $d\geq t$, in which case $\gamma(t,d)=t!$, and by Rains~\cite{Rains98} in general. It is the number of 
permutations $\sigma\in\S_t$ (the symmetric group~\cite{Sagan01}) such that $(\sigma(1),\sigma(2),\dots,\sigma(t))$ has no increasing subsequence of length greater than $d$. 
Thus, for example, $\gamma(1,d)=1$ and $\gamma(2,d)=2$ for all $d\geq 2$.

A unitary 1-design must satisfy
\begin{equation}\label{eq:1design}
\sum_{x\in\mathscr{D}} w(x)\, U(x)\otimes U(x)^\dag \;=\; \int_{\PU(d)}\d\muu(x)\,U(x)\otimes U(x)^\dag \;=\;  \frac{1}{d}\,T \;,
\end{equation}
where the RHS of Eq.~(\ref{eq:wdesign}) is now explicitly evaluated (simply consider a matrix component of the integral 
in the standard product basis and use Schur's lemma). Since $T$ has eigenvalues of $1$ and $-1$, respectively, on the 
symmetric and antisymmetric subspaces of $\C^d\otimes\C^d$, and thus $\rank(T)=d^2$, we must have $|\mathscr{D}|\geq d^2$ 
with equality only if $\mathscr{D}$ is an (orthogonal) unitary operator basis, i.e. $\tr[U(x)^\dag U(y)]=0$ for all $x\neq y\in\mathscr{D}$, 
and $w(x)=1/|\mathscr{D}|$. This fact is more apparent when Eq.~(\ref{eq:1design}) is rewritten in terms of superoperators:
\begin{equation}
\sum_{x\in\mathscr{D}} w(x)\, \KetBra{U(x)} \;=\; \frac{1}{d}\,\Is \;.
\end{equation}
In this form it is clear that unitary 1-designs are equivalent to tight unitary frames~\cite{Horn05,Scott06}. 
The unitary operators with matrix elements~\cite{Horn05}
\begin{equation}\label{UTOF}
\bra{j}U_m\ket{k}\;\coloneq\; \frac{1}{\sqrt{d}}\exp\left[\frac{2\pi ijk}{d}+\frac{2\pi i(j+kd)m}{n}\right] \;, 
\end{equation}
for $m=0,\dots,n-1$, provide explicit examples of (unweighted) unitary 1-designs for all $n=|\mathscr{D}|\geq d^2$. 

To treat the general case, for an arbitrary permutation $\sigma\in \S_n$, define the {\em permutation operator}  
\begin{equation}\label{eq:perm}
P(\sigma) \;=\; P_{\sigma(1)\sigma(2)\dots \sigma(n)} \;\coloneq\; \sum_{j_1,j_2,\dots,j_n} \ket{j_1}\bra{j_{\sigma(1)}}\otimes\ket{j_2}\bra{j_{\sigma(2)}}\otimes \dots\otimes \ket{j_n}\bra{j_{\sigma(n)}}\;,
\end{equation}
which acts on $(\C^d)^{\otimes n}$ by permuting its subsystems accordingly: 
\begin{equation}
P_{k_1 k_2\dots k_n}\ket{\psi_{k_1}}\otimes\ket{\psi_{k_2}}\otimes\dots\otimes\ket{\psi_{k_n}} \;=\;\ket{\psi_1}\otimes\ket{\psi_2}\otimes\dots\otimes\ket{\psi_n}\;, 
\end{equation}
and in terms of operators,
\begin{equation}\label{eq:permop}
P_{k_1 k_2\dots k_n} \big( A_{k_1}\otimes A_{k_2} \otimes\dots\otimes A_{k_n} \big)\, {P_{k_1 k_2\dots k_n}}^\dag \;=\;A_1\otimes A_2\otimes\dots\otimes A_n \;.
\end{equation}
The composition of permutation operators then follows that for permutations, $P(\sigma)P(\tau)=P(\sigma\tau)$, which means 
${P(\sigma)}^\dag=P(\sigma^{-1})$. Note that $P_{21}=T$.

In general, the RHS of Eq.~(\ref{eq:wdesign}) can be integrated explicitly using group theoretical methods~\cite{Collins03,Collins06}. 
The result is
\begin{equation}
\int_{\PU(d)}\d\muu(x)\,U(x)^{\otimes t}\otimes \big(U(x)^{\otimes t}\big)^\dag \;=\; \sum_{\sigma,\tau\in \S_t} \Wg(d,t,\sigma\tau^{-1}) P_{\tau(1)+t,\dots,\tau(t)+t,\sigma^{-1}(1),\dots,\sigma^{-1}(t)}\;,
\end{equation}
where 
\begin{equation}
\Wg(d,t,\sigma) \;\coloneq\;\frac{1}{t!^2}\mathop{\sum_{\lambda\vdash t}}_{l(\lambda)\leq d}\frac{\chi^\lambda(1)^2\chi^\lambda(\sigma)}{s_{\lambda,d}(1)}\;,
\end{equation}
is called the {\em Weingarten function}. Here the sum is over all partitions 
$\lambda=(\lambda_1,\dots,\lambda_t)$ of the integer $t$ (i.e. nonincreasing sequences of nonnegative integers 
summing to $t$) with length $l(\lambda)\leq d$, where $l(\lambda)\coloneq\max_{\lambda_j>0}\,j$. The 
character on the conjugacy class $\mathrm{K}_\lambda$ of $\S_t$ corresponding to $\lambda\vdash t$ is denoted by $\chi^\lambda$ and 
we take $s_{\lambda,d}(1)=s_{\lambda,d}(1,\dots,1)$ for the Schur function $s_{\lambda,d}(x_1,\dots,x_d)$~\cite{Sagan01}. 
For any partition $\lambda\vdash t$ one has
\begin{equation}
s_{\lambda,d}(1) \;=\; \frac{1}{t!}\sum_{\mu\vdash t} d^{l(\mu)}\chi^\lambda(\mu)|\mathrm{K}_\mu| \;,
\end{equation}
and in particular, $s_{(1,1),d}(1)=d(d+1)/2$ and $s_{(2,0),d}(1)=d(d-1)/2$. This means $\Wg(d,2,(1,1))=1/(d^2-1)$ and
$\Wg(d,2,(2,0))=-1/d(d^2-1)$, giving
\begin{align}\label{eq:int2des}
\int_{\PU(d)}\!\!\d\muu(x)\, U(x) \otimes U(x)& \otimes U(x)^\dag \otimes U(x)^\dag \nonumber\\ 
&\;=\;  \frac{1}{d^2-1}\left(P_{3412}+P_{4321}\right) - \frac{1}{d(d^2-1)}\left(P_{4312}+P_{3421}\right) \;.
\end{align}
Thus our definition of a unitary 2-design [Eq.~(\ref{eq:wdesign})] can be rewritten as
\begin{align}\label{eq:def2des}
\sum_{x\in\mathscr{D}}w(x)\, U(x) \otimes U(x)& \otimes U(x)^\dag \otimes U(x)^\dag \nonumber\\ 
&\qquad\!\! =\;  \frac{1}{d^2-1}\left(P_{3412}+P_{4321}\right) - \frac{1}{d(d^2-1)}\left(P_{4312}+P_{3421}\right) \;.
\end{align}
The following is partly due to Gross~{\it et al}.~\cite[Theorem 2]{Gross06}.

\begin{thm}\label{thm:2design}
Let\/ $(\mathscr{D},w)$, $\mathscr{D}\subset\PU(d)$, be a weighted\/ $2$-design. Then\/ 
\begin{equation}\label{eq:designbound}
|\mathscr{D}| \;\geq\; (d^2-1)^2+1 \;,
\end{equation} 
with equality only if\/ $w(x)= 1/|\mathscr{D}|$ and 
\begin{equation}\label{eq:tight2designcondition}
|\tr[U(x)^\dag U(y)]|^2 \;=\; 1-\frac{1}{d^2-1} \;,
\end{equation}
for all\/ $x\neq y\in\mathscr{D}$. 
\end{thm}

\begin{proof}
Multiplying Eq.~(\ref{eq:def2des}) on the left by $P_{2341}$ and on the right by ${P_{2341}}^{\dag}=P_{4123}$ implies 
\begin{align}
\sum_{x\in\mathscr{D}}w(x)\, U(x)^\dag \otimes U(x)& \otimes U(x) \otimes U(x)^\dag \nonumber\\ 
&\;=\; \frac{1}{d^2-1}\left(P_{3412}+P_{2143}\right) - \frac{1}{d(d^2-1)}\left(P_{3142}+P_{2413}\right) \;,
\end{align}
given Eq.~(\ref{eq:permop}) and since, for example, $P_{2341}P_{3412}P_{4123}=P_{2341}P_{2341}=P_{3412}$.
Now multiply this equation on the right by $A\otimes I\otimes I$, where $A\in\End(\C^d)\otimes\End(\C^d)$,
and trace out the first pair of subsystems. The result can be written in terms of a superoperator:
\begin{align}
\mathcal{S}(A) &\;\coloneq\; \sum_{x\in\mathscr{D}}w(x)\, \tr\big[\big(U(x)^\dag\otimes U(x)\big)A\big]\, U(x)\otimes U(x)^\dag  \\
&\;=\;\frac{A+\tr(AT)T}{d^2-1}\,-\,\frac{\big(I\otimes\tr_1(AT)\big)T+\big(\tr_2(AT)\otimes I\big) T}{d(d^2-1)} \;,
\end{align}
by rewriting the permutation operators explicitly in terms of their definition [Eq.~(\ref{eq:perm})] and simplifying.

Now let $\{E_k\}_{k=0}^{d^2-1}$ be an orthonormal operator basis for $\End(\C^d)$ with the choice $E_0=I/\sqrt{d}$. 
The remaining operators $E_1,\dots,E_{d^2-1}$ then span the $(d^2-1)$-dimensional subspace of traceless operators: 
$\tr(E_k)=\sqrt{d}\tr(E_0^\dag E_k)=0$ for all $k>0$. Consider the action of $\mathcal{S}$ on $(E_j\otimes E_k)T\,$:
\begin{align}
\mathcal{S}\big((E_j\otimes E_k)T\big) &\;=\; \frac{1}{d^2-1} \Big( (E_j\otimes E_k)T + d^2\delta_{j0}\delta_{k0}(E_0\otimes E_0)T \nonumber\\ 
&\qquad\qquad\qquad\qquad\qquad\qquad -\delta_{j0}(E_0\otimes E_k)T-\delta_{0k}(E_j\otimes E_0)T \Big)\\
&\;=\; \begin{cases}
(E_0\otimes E_0)T\,, &\; j=k=0\,; \\ 
(E_j\otimes E_k)T/(d^2-1)\,, &\; j,k>0\,; \\
0\,, &\; \text{otherwise}\,,
\end{cases}
\end{align}
identifying $I=\sqrt{d}E_0$. Thus the $d^4$ orthonormal operators $(E_j\otimes E_k)T$ diagonalize $\mathcal{S}$ and, 
in particular, $\rank'(\mathcal{S})=(d^2-1)^2+1$ (ordinary rank). But we must have 
$|\mathscr{D}|\geq\rank'(\mathcal{S})$, which is Eq.~(\ref{eq:designbound}). 

If $|\mathscr{D}|=\rank'(\mathcal{S})$ then $\{U(x) \otimes U(x)^\dag\}_{x\in\mathscr{D}}$ is necessarily a 
linearly independent set. Fixing $y\in\mathscr{D}$ and considering $\mathcal{S}(U(y)\otimes U(y)^\dag)$ shows that
\begin{equation}
(d^2-1)\sum_{x\in\mathscr{D}}w(x)\, |\tr[U(x)^\dag U(y)]|^2\, U(x) \otimes U(x)^\dag \;=\; U(y) \otimes U(y)^\dag +\Big(d-\frac{2}{d}\Big)\,T \;,
\end{equation}
which, upon setting $T=d\sum_{x\in\mathscr{D}} w(x) U(x)\otimes U(x)^\dag$ [Eq.~(\ref{eq:1design})], can be rewritten as
\begin{align}
\big\{\big((d^2-1)^2+&1\big)w(y)-1\big\}\,U(y) \otimes U(y)^\dag \nonumber\\ 
&+\sum_{x\neq y}w(x)\big\{(d^2-1)|\tr[U(x)^\dag U(y)]|^2-d^2+2\big\}\, U(x) \otimes U(x)^\dag \;=\; 0 \;.
\end{align}
When $|\mathscr{D}|=(d^2-1)^2+1$ linear independence thus requires $(d^2-1)|\tr[U(x)^\dag U(y)]|^2=d^2-2$ for all 
$x\neq y$ and $\big((d^2-1)^2+1\big)w(y)=|\mathscr{D}|w(y)=1$. The same is true for all $y\in\mathscr{D}$.
\end{proof}

In general, for each positive integer $t$ and $d$, we would like to know the quantity $N(t,d)$, which we use to denote 
the minimum number of unitaries needed to construct a weighted $t$-design in $\PU(d)$, or less ambitiously, bounds on 
this quantity. This is a difficult problem. A general lower bound, however, might be obtainable from the theory of 
Levenshtein~\cite{Levenshtein98a,Levenshtein98b}.

Our own numerical searches have not revealed the existence of $2$-designs achieving $|\mathscr{D}|=(d^2-1)^2+1$ 
and Gross~{\it et al}.~\cite{Gross06} have conjectured their nonexistence. If such designs did exist then they might 
be dubbed {\em tight\/} designs, which is standard terminology in the theory of $t$-designs~\cite{Delsarte77}
(but unrelated to the concept of tight frames). As noted in Theorem~\ref{thm:2design}, tight 
$\PU(d)$ $2$-designs are necessarily equiangular. They are analogous to tight $\C P^{d-1}$ $2$-designs, i.e.
symmetric informationally complete POVMs (SIC-POVMs)~\cite{Renes04}, which in contrast are conjectured to exist in all dimensions.
The analogy to a complete family of mutually unbiased bases (MUBs)~\cite{Ivanovic81,Wootters89}, however, does exist in certain
dimensions. 

A subset $\{U_j\}_{j=0}^{d^2-1}\subset\U(d)$ is a unitary operator basis for $\End(\C^d)$ if $\tr({U_j}^\dag U_k)=d\delta_{jk}$ 
for all $0\leq j,k \leq d^2-1$.  In analogy with the case of vector bases, we call a pair of unitary operator bases, 
$\{U_j\}_{j=0}^{d^2-1}$ and $\{V_k\}_{k=0}^{d^2-1}$, {\em mutually unbiased\/} if
\begin{equation}\label{eq:MUUBsdef}
|\tr({U_j}^\dag V_k)|^2 \;=\; 1 
\end{equation}
for all $0\leq j,k\leq d^2-1$. Define the embedding $\vartheta:\U(d)\hookrightarrow\Houc\cong\R^{(d^2-1)^2}$ by 
\begin{equation}
\Ket{\vartheta(U)} \;\coloneq\; \Ptr\Ket{\ketbra{U}} \;=\; \Ket{\ketbra{U}-I/d^2} \;,
\end{equation}
where $\ketbra{U}$, $\Ptr$ and $\Houc$ are defined in Eqs.~(\ref{eq:ketU}), (\ref{eq:proj0})
and (\ref{eq:Houc}), respectively. The set $\{\vartheta(U_j)\}_{j=0}^{d^2-1}$ then specifies the vertices of a regular 
simplex in the $(d^2-1)$-dimensional subspace of $\Houc$ for which its members span.
Mutually unbiased bases correspond to orthogonal subspaces,
\begin{equation}
\BraKet{\vartheta(U_j)}{\vartheta(V_k)} \;=\; |\braket{U_j}{V_k}|^2-\frac{1}{d^2} \;=\; \frac{1}{d^2}|\tr({U_j}^\dag V_k)|^2-\frac{1}{d^2} \;=\;0\;,
\end{equation} 
of which, there can be at most $(\dim\Houc)/(d^2-1)=d^2-1$ many. A set of $d^2-1$ unitary operator 
bases with the property that each pair is mutually unbiased is thus called a {\em complete\/} set of mutually 
unbiased unitary-operator bases (MUUBs).

Now consider an arbitrary family of subsets, $\mathscr{B}_0,\dots,\mathscr{B}_{m-1}\subset\PU(d)$, where each member
$\mathscr{B}_a=\{e_j^a\}_{j=0}^{d^2-1}$ specifies a unitary operator basis $\{U(e_j^a)\}_{j=0}^{d^2-1}$ 
and is appointed a positive weight $w_a$. By Theorem~\ref{thm:welch}, if
\begin{equation}\label{eq:wbasesdef}
\sum_{a,b=0}^{m-1}w_a w_b\sum_{j,k=0}^{d^2-1}\,|\tr[U(e_j^a)^\dag U(e_k^b)]|^{4} \;=\; 2 \;, 
\end{equation}
then their union $\mathscr{D}=\cup_a\mathscr{B}_a$ forms a weighted $2$-design with weight $w(x)=\sum_a w_a 1_{\mathscr{B}_a}(x)$. 
In the context of quantum process tomography it is desirable for the weight to remain constant across elements of the 
same basis (so the POVM that the design specifies can be implemented by a series of orthogonal measurements). 
We have thus made this a requirement. The set indicator function, $1_\mathscr{S}(x)\coloneq 1$ if $x\in\mathscr{S}$ 
and 0 otherwise, is used to take care of any multiplicity across different bases. Notice that the normalization of $w(x)$ 
implies normalization of the basis weights: $\sum_a w_a=1/d^2$.

It is straightforward to confirm [via Eq.~(\ref{eq:wbasesdef})] that a complete set of MUUBs forms a unitary 2-design 
when $w_a=1/md^2$. The following theorem shows that such sets are optimal, in that we always need $m\geq d^2-1$ unitary 
operator bases to construct a weighted 2-design, with equality only if the bases are mutually unbiased.

\begin{thm}\label{thm:MUUBs}
Let $d>1$ and let\/ $\mathscr{B}_0,\dots,\mathscr{B}_{m-1}\subset\PU(d)$ specify a family of unitary operator bases for\/ 
$\End(\C^d)$, where the union\/ $\mathscr{D}=\cup_a\mathscr{B}_a$ forms a weighted\/ $2$-design with weight function\/ 
$w(x)=\sum_a w_a 1_{\mathscr{B}_a}(x)$ for some choice of basis weights\/ $w_0,\dots,w_{m-1}>0$. Then\/ $m\geq d^2-1$ with 
equality only if\/ $w_a=1/md^2$ for all\/ $a$ and the bases are pairwise mutually unbiased. 
\end{thm}
\begin{proof}
Theorem~\ref{thm:2design} with $|\mathscr{D}|=md^2$ immediately shows that we require $m\geq d^2-2+2/d^2$, which means 
$m\geq d^2-1$ whenever $d>1$. 
In the case of equality, note that by Theorem~\ref{thm:welch} [or Eq.~(\ref{eq:wbasesdef})] we require 
\begin{equation}
d^6 \sum_a {w_a}^2 + \sum_{a \neq b}w_aw_b \sum_{j,k} (\lambda_{jk}^{ab})^2 \;=\; 2 \;, \label{eq:thmMUBsproofa} 
\end{equation}
where we have defined the positive numbers $\lambda_{jk}^{ab}\coloneq|\tr[U(e^a_j)^\dag U(e^b_k)]|^2$. Moreover, Theorem~\ref{thm:welch} 
implies that the LHS of Eq.~(\ref{eq:thmMUBsproofa}) is minimal with respect to the variables $w_a$ and $\lambda_{jk}^{ab}$ 
under the appropriate constraints, two of which are $\sum_a w_a = 1/d^2$ and 
\begin{equation}\label{eq:thmMUBsproofb}
\sum_{j,k}\lambda_{jk}^{ab}\;=\;\sum_{j,k}|\BraKet{U(e^a_j)}{U(e^b_k)}|^2\;=\;d^2\Tr(\Is \cdot \Is) \;=\; d^4 \;,
\end{equation}
since all unitary operator bases satisfy $\sum_j\KetBra{U(e_j)}=d\Is$.
We will now minimize the LHS of Eq.~(\ref{eq:thmMUBsproofa}) under these two constraints. 
The minimum of $\sum_{j,k} (\lambda_{jk}^{ab})^2$ subject to Eq.~(\ref{eq:thmMUBsproofb}) occurs only when $\lambda_{jk}^{ab}=1$ for all 
$0\leq j,k\leq d^2-1$, i.e., when $\mathscr{B}_a$ and $\mathscr{B}_b$ are mutually unbiased. 
Then the LHS of Eq.~(\ref{eq:thmMUBsproofa}) reduces to
\begin{equation}\label{eq:thmMUBsproofc}
d^6 \sum_a w_a^2 + d^4\sum_{a \neq b}w_aw_b \;=\; d^4(d^2-1) \sum_a w_a^2 + 1 \;,
\end{equation}
and here the minimum (under $\sum_a w_a = 1/d^2$) occurs only when $w_a = 1/md^2$ for all $0 \leq a\leq m-1$.
With this value, Eq.~(\ref{eq:thmMUBsproofc}) reduces to the RHS of Eq.~(\ref{eq:thmMUBsproofa}) when $m=d^2-1$. Equality 
in Eq.~(\ref{eq:thmMUBsproofa}) thus requires the bases to be pairwise mutually unbiased and $w_a = 1/md$ whenever $m=d^2-1$. 
\end{proof}

Theorem~\ref{thm:MUUBs} is the equivalent of Ref.~\cite[Theorem~3.2]{Roy07} for the case of unitary designs.
Many examples of unweighted unitary $2$-designs were described by Gross~{\it et al}.~\cite{Gross06}. Of these, 
the {\em Clifford designs\/} were found closest to optimal. These are sets of unitary operator bases which
form subgroups of the projective Clifford group $\PC(d)$~\cite{Appleby05,Flammia06} and have cardinalities $|\mathscr{D}|=kd^2(d^2-1)$ for some integer $k$. 
When $k=1$, Clifford designs are known to exist in dimensions $d=2,3,5,7,11$~\cite{Chau05}. By Theorem~\ref{thm:MUUBs}, each of 
these examples must be the union of a complete set of MUUBs. Although no unweighted unitary $2$-designs of smaller size were found
by Gross~{\it et al}.~\cite{Gross06}, weighted unitary $2$-designs can surpass this record. This is the case for $\PU(2)$ $2$-designs, 
which are described in detail next.

\subsection{PU(2) $\bm{t}$-designs}
\label{sec:qubitdesigns}

In dimension 2, unitary designs are equivalent to real projective designs, which in turn are 
equivalent to antipodal spherical designs. To see this, simply note that $\PU(2)\cong\R P^3$ through the relation
\begin{equation}\label{eq:isomorph}
e^{i\phi}U \;=\; r_0 I + i(r_1 X + r_2 Y + r_3 Z) \;,
\end{equation}
where $X\coloneq\big(\begin{smallmatrix} 0 & 1 \\ 1 & 0\end{smallmatrix}\big)$, 
$Y\coloneq\big(\begin{smallmatrix} 0 & -i \\ i & 0\end{smallmatrix}\big)$, and 
$Z\coloneq\big(\begin{smallmatrix} 1 & 0 \\ 0 & -1\end{smallmatrix}\big)$ are the Pauli matrices. Each unit vector $(r_0,r_1,r_2,r_3)\in\R^4$ specifies a line 
in $\R P^3$, and through Eq.~(\ref{eq:isomorph}), an equivalence class of unitaries $U\in\U(2)$ differing only be a phase 
factor. Under this map each $t$-design in $\PU(2)$ gives a $t$-design in $\R P^3$ and vice versa. This is because 
distances are preserved: $|\tr(U^\dag V)|^2=4\braket{r}{s}^2$ where $\braket{r}{s}\coloneq\sum_k r_k s_k$ and $r$ 
(respectively, $s$) corresponds to $U$ (respectively, $V$) through Eq.~(\ref{eq:isomorph}). Theorem~\ref{thm:welch} 
then transforms to the equivalent for real projective designs. This relationship also gives 
\begin{equation}
\gamma(t,2) \;=\; \frac{(2t)!}{t!(t+1)!} \;. 
\end{equation}
Real projective designs are rarely studied in the literature. It is well known, however, that $t$-designs in 
$\R P^{n-1}$ are equivalent to antipodal $(2t+1)$-designs in $S^{n-1}$ with twice as many points (assuming antipodal 
pairs are appointed the same weight). The antipodal points of the spherical design are simply the intersections 
between the lines of the real projective design and the unit sphere. Additionally, an antipodal spherical 
$(2t+1)$-design can be created from $(2t)$-design by simply appending the antipodal points to the design:
if $\mathscr{D}$ is a $(2t)$-design in $S^{n-1}$ then $\mathscr{D}\cup(-\mathscr{D})$ is an antipodal $(2t+1)$-design 
in $S^{n-1}$.

\begin{table}
\begin{tabular}{|c|c|c|c|}\hline 
$\quad t\quad$ & $\;$ Delsarte $\;$ & $\quad N(t,2)\geq\quad$ & $\quad N(t,2)\leq\quad$ \\\hline
2 & 10 & 11\footnotemark[1] (12\footnotemark[2]) & 11\footnotemark[3]  (12\footnotemark[4]) \\
3 & 20 & 21\footnotemark[1] & 23\footnotemark[5] (24\footnotemark[6]) \\
4 & 35 & 37\footnotemark[7] & 43\footnotemark[5]  \\
5 & 56 & 60\footnotemark[7] &  60\footnotemark[8] \\
6 & 84 & 85\footnotemark[1] (89\footnotemark[9]) &  \\
7 & 120 & 134\footnotemark[7]  & 264\footnotemark[10]  \\
8 & 165 & 166\footnotemark[1]  (180\footnotemark[9])  & \\
9 & 220 & 250\footnotemark[7]  & 360\footnotemark[11]  \\
10 & 286 & 287\footnotemark[1]  (318\footnotemark[9])  & \\\hline
\end{tabular}
\footnotetext[1]{No tight $S^3$ $(2t+1)$-designs exist for $t>1$~\cite{Delsarte77,Bannai79,Bannai80}.}
\footnotetext[2]{No antipodal unweighted 22-point $S^3$ 5-designs exist~\cite{Reznick95}.}
\footnotetext[3]{A weighted 11-point $\PU(2)$ 2-design exists [Eq~(\ref{eq:w22des})].}
\footnotetext[4]{The 24 vertices of the 24-cell form an antipodal unweighted $S^3$ 5-design~\cite{Sloane03}, and thus 
also an unweighted $\PU(2)$ 2-design. This design is a minimal subgroup of $\PC(2)$.}
\footnotetext[5]{A weighted 23-point $S^3$ 6-design and a weighted 43-point $S^3$ 8-design exist~\cite{Hardin94}.}
\footnotetext[6]{The projective Clifford group $\PC(2)$ is an unweighted $\PU(2)$ 3-design. The corresponding $S^3$ 7-design is formed by the vertices of 2 copies of the 24-cell~\cite{Sloane03}.}
\footnotetext[7]{The linear programming bounds for weighted $S^3$ $(2t+1)$-designs of Ref.~\cite{Harpe06}.}
\footnotetext[8]{The 120 vertices of the 600-cell form an antipodal unweighted $S^3$ 11-design. This is the unique minimal unweighted $S^3$ 11-design~\cite{Boyvalenkov01,Andreev00}.}
\footnotetext[9]{Yudin's bound~\cite{Yudin97} on unweighted spherical designs gives $|\mathscr{D}|\geq \pi/\big(\pi-2x\sqrt{1-x^2}-2\arcsin x\big)$ for unweighted $\PU(2)$ $t$-designs, where $x$ is the largest zero of the Jacobi polynomial $P^{(3/2,3/2)}_{2t+1}(x)$.}
\footnotetext[10]{An antipodal weighted 528-point $S^3$ 15-design can be constructed from shells of a Euclidean lattice~\cite{Harpe06}.}
\footnotetext[11]{The union of the 120 vertices and the 600 face centers of the 600-cell form a weighted antipodal $S^3$ 19-design~\cite{Salihov75}. The vertices have weight $1/504$ and the faces have weight $2/1575$.}
\caption{Known bounds on $N(t,2)$, the minimum cardinality of a weighted $t$-design in $\PU(2)$. The Delsarte lower bound 
[Eq.~(\ref{eq:designbound2})] is included as a reference point, and for completeness, bounds on the minimum cardinality
for an unweighted design are included in parentheses.}
\label{tbl:dim2tdes}
\end{table}

The above relationships can be used to translate known results in the literature to the case of unitary designs. 
For example, the lower bound of Delsarte {\it et al.}~\cite{Delsarte77} on the number of points needed 
to construct a $(2t+1)$-design in $S^3$ shows that 
\begin{equation}\label{eq:designbound2}
|\mathscr{D}|\;\geq\;\frac{1}{6}(t+1)(t+2)(t+3)
\end{equation}
for a $t$-design in $\PU(2)$, with equality only if the design is unweighted~\cite{Levenshtein98b}, i.e. $w(x)= 1/|\mathscr{D}|$.
A design which achieves this bound is generally called {\em tight\/}. It is known, however, that tight 
$S^3$ $(2t+1)$-designs exist only for the trivial $t=1$ case~\cite{Delsarte77,Bannai79,Bannai80} (see 
Ref.~\cite{Bannai04} for a summary). Thus we can increase the RHS of Eq.~(\ref{eq:designbound2}) by 1 when $t>1$.
Further bounds on the cardinality of a $\PU(2)$ $t$-design are summarized in Table~\ref{tbl:dim2tdes}. 

The first row of Table~\ref{tbl:dim2tdes} corresponds to $t=2$, which we now explain in detail. The Delsarte bound [Eq.~(\ref{eq:designbound2})] shows 
that $\PU(2)$ $2$-designs must have at least 10 points. However we can increase this bound to 11 since there are no tight 
$\PU(2)$ $2$-designs. The 11 columns of the matrix 
\begin{equation}\label{eq:w22des}
\left[\begin{array}{rrrrrrrrrrr}
1 &\phantom{-} 0 &\phantom{-} 0 &\phantom{-} 0 &\phantom{-} 0 &\phantom{-} a &\phantom{-} a &\phantom{-} a &\phantom{-} a &\phantom{-} a &\phantom{-} a \\
0 & a & -a & a & a & b & -b & 0 & 0 & 0 & 0 \\
0 & a & a & -a & a & 0 & 0 & b & -b & 0 & 0 \\
0 & a & a & a & -a & 0 & 0 & 0 & 0 & b & -b 
\end{array}\right]\;,
\end{equation}
where $a=1/\sqrt{3}$ and $b=\sqrt{2/3}$, specify a weighted $\PU(2)$ $2$-design [through Eq.~(\ref{eq:isomorph})] 
which achieves the bound. The weight appointed to the first column is $1/16$ while the remaining 
all have weight $3/32$. Reznick~\cite{Reznick95} has shown that there are no antipodal unweighted $S^3$ 5-designs with 22 
points. Thus this design is necessarily weighted. The 24 vertices of the 24-cell form an antipodal unweighted $S^3$ 
5-design~\cite{Sloane03}, and thus also an unweighted $\PU(2)$ 2-design with 12 points. The elements of this design correspond 
to the subgroup $\langle HR,R^2\rangle$ of the Clifford group $\mathrm{C}(2)=\langle H,R \rangle$, where 
$H=\frac{1}{\sqrt{2}}\big(\begin{smallmatrix} 1 & \phantom{-}1 \\ 1 & -1 \end{smallmatrix}\big)$ and 
$R=\big(\begin{smallmatrix} 1 & 0 \\ 0 & i\end{smallmatrix}\big)$. It is formed by the union of a 
complete set of MUUBs and is the first in the family of Clifford designs.

Finally, Shamsiev's explicit constructions of antipodal weighted spherical designs~\cite[Theorem 1]{Shamsiev06} show that 
weighted $\PU(2)$ $t$-designs with $(t+1)^3/2$ points exist for all odd $t$. This upper bound on the cardinality together 
with the Delsarte lower bound [Eq.~(\ref{eq:designbound2})] means that $N(t,2)=\Theta(t^3)$.

\section{Optimal ancilla-assisted quantum process tomography}
\label{sec:optimalprocess}

We now return to our immediate task of optimizing the measurements used for ancilla-assisted quantum process tomography.
Throughout this section we assume that either $\Qs=\Qu(\C^d\otimes\C^d)$, $\GC$ or $\UC$. The results of 
Sec.~\ref{sec:optimal} are first summarized for these specific cases. 

Recall that our error for tomographic reconstructions of quantum states was defined in terms of the 
Hilbert-Schmidt distance [Eq.~(\ref{eq:error})]:
\begin{equation}
e^{(F,Q)}(\rho) \;\coloneq\; \mathop{\textrm{\large E}}_{y_1,\dots,y_N}\Big[\,\|\rho-\hat{\rho}(y_1,\dots,y_N)\|^2\,\Big] \;,
\end{equation}
where $\hat{\rho}$ is the linear tomographic estimate of $\rho$ given $N$ measurement outcomes $y_1,\dots,y_N$ 
[Eqs.~(\ref{pestimator}) and (\ref{linearreconstructd})]. When applied to the output states of quantum channels, 
with fixed input $\rho_\mathrm{i}=\ketbra{I}$, this distance measure naturally induces the analogous Hilbert-Schmidt distance 
for superoperators (see the appendix):
\begin{equation}
\|\mathcal{E}-\hat{\mathcal{E}}\| \;=\; \|\rho-\hat{\rho}\| \;,
\end{equation}
where $\rho=(\mathcal{E}\otimes\mathcal{I})(\rho_\mathrm{i})$ and $\hat{\rho}=(\hat{\mathcal{E}}\otimes\mathcal{I})(\rho_\mathrm{i})$ 
through the Jamio{\l}kowski isomorphism [Eq.~(\ref{eq:Jam})], and $\|\mathcal{S}\|\coloneq\sqrt{\Tr(\mathcal{S}^\dag\mathcal{S})}$ 
for any superoperator $\mathcal{S}$. Although there are more appropriate distance measures for quantum channels, which 
properly reflect the probabilistic interpretation of a quantum state, the Hilbert-Schmidt distance is the most 
natural choice for linear tomographic reconstructions of quantum states.

Now setting $\rho=\rho(\sigma,U)\coloneq U\sigma U^\dag$ for some fixed output state $\sigma\in\Qs$, recall that we defined  
the average [Eq.~(\ref{eq:averror})] and worst-case [Eq.~(\ref{estwcineq})] error over different Hilbert-space 
orientations $U_\mathrm{s},U_\mathrm{a}\in\U(d)\,$:
\begin{align}
e_\mathrm{av}^{(F,Q)}(\sigma) &\;\coloneq\; \iint_{\U(d)}\d\muu(U_\mathrm{s})\d\muu(U_\mathrm{a})\:e^{(F,Q)}(\rho(\sigma,U_\mathrm{s}\otimes U_\mathrm{a}))\;; \\
e_\mathrm{wc}^{(F,Q)}(\sigma) &\;\coloneq\; \sup_{U_\mathrm{s},U_\mathrm{a}\in\U(d)}\:e^{(F,Q)}(\rho(\sigma,U_\mathrm{s}\otimes U_\mathrm{a}))\;. 
\end{align}
Finally, recalling that the subsets $\GC$ and $\UC$ respectively span $\delta'=d^2(d^2-1)+1$ and $\delta'=(d^2-1)^2+1$ dimensions of 
$\H(\C^d\otimes\C^d)$, the following corollary restates Theorem~\ref{thmest} and Corollary~\ref{corest} for 
these special cases of interest.

\begin{cor}\label{cor:process}
Let\/ $F:\mathfrak{B}(\mathscr{X})\rightarrow\H(\C^d\otimes\C^d)$ be a POVM which is 
informationally complete w.r.t.\ $\Qs\subseteq\Qu(\C^d\otimes\C^d)$. 
Then for any fixed quantum state\/ $\sigma\in\Qs$,
\begin{equation}\label{eq:errorrates}
e_\mathrm{wc}^{(F,Q)}(\sigma) \;\geq\; e_\mathrm{av}^{(F,Q)}(\sigma) \;\geq\;\begin{cases}
\displaystyle\phantom{\bigg\|}\frac{1}{N}\Big(d^4 + d^2 - 1 - \tr(\sigma^2) \Big)\,, &\; \text{if\/ $\Qs=\Qu(\C^d\otimes\C^d)$}\,; \\ 
\displaystyle\phantom{\bigg\|}\frac{1}{N}\Big(d^4 - d^2 + 1/d^2 - \tr(\sigma^2) \Big)\,, &\; \text{if\/ $\Qs=\GC$}\,; \\ 
\displaystyle\phantom{\bigg\|}\frac{1}{N}\Big(d^4 - 3d^2 + 3 - \tr(\sigma^2) \Big)\,, &\; \text{if\/ $\Qs=\UC$}\,,
\end{cases} 
\end{equation}
for all reconstruction OVDs\/ $Q$. Furthermore, equality in the RHS of Eq.~(\ref{eq:errorrates}) occurs if and only if\/ $Q=R$ and $F$ is a tight rank-one 
POVM with\/ $\supp(\Fs)=\Span(\Qs)$, in which case we also have equality in the LHS of Eq.~(\ref{eq:errorrates}). 
\end{cor}

Tight rank-one POVMs thus describe the class of optimal measurements for {\em linear\/} ancilla-assisted quantum process tomography 
in both an average and worst-case sense. But do such measurements exist? When $\Qs=\Qu(\C^d\otimes\C^d)$, which is the class of 
output states of general quantum operations (and included in Corollary~\ref{cor:process} for comparison), we recover the results 
of Ref.~\cite{Scott06}. Here it was found that tight rank-one POVMs exist and are in fact equivalent to weighted complex 
projective 2-designs.

Now consider the case $\Qs=\UC$. Our condition for a tight rank-one POVM [Eq.~(\ref{tightdfneq3})] with 
$D=d^2$, $\delta=(d^2-1)^2+1$ and  
\begin{equation}
\Ps_{\Fs} \;=\; \Ps_{\UC} \;=\; \KetBra{\lambda_0\otimes\lambda_0}+\sum_{j,k>0}\KetBra{\lambda_j\otimes\lambda_k} \;,
\end{equation}
becomes
\begin{equation}\label{eq:ucsuperop}
\Fs \;=\;\int_\mathscr{X} \d\tau(x)\,\KetBra{P(x)} \;=\; \KetBra{\lambda_0\otimes\lambda_0}+\frac{1}{d^2-1}\sum_{j,k>0}\KetBra{\lambda_j\otimes\lambda_k}\;,
\end{equation}
using the orthonormal Hermitian operator basis $\{\lambda_k\}_{k=0}^{d^2-1}$ of Sec.~\ref{sec:process} 
[see above Eq.~(\ref{eq:hermopbasis})]. Equivalently, under the isomorphism $\Ket{A}\Bra{B}\leftrightarrow A\otimes B^\dag$ 
we can rewrite this last form as
\begin{equation}
\int_\mathscr{X} \d\tau(x)\, P(x) \otimes P(x) \;=\; (\lambda_0\otimes\lambda_0)\otimes(\lambda_0\otimes\lambda_0)+\frac{1}{d^2-1}\sum_{j,k>0}(\lambda_j\otimes\lambda_k)\otimes(\lambda_j\otimes\lambda_k)\;.
\end{equation}
Now multiplying on the left by $P_{1432}$, taking the trace, and applying the easily confirmed identity 
$\tr[P_{1432}(A\otimes B\otimes C\otimes D)]=\tr(A)\tr(C)\tr(BD)$, we find that the types of tight rank-one POVMs 
corresponding to $\UC$ must satisfy
\begin{equation}\label{eq:tightUC1}
\int_\mathscr{X} \d\tau(x)\, \tr_\mathrm{s}\!\big[\{\tr_\mathrm{a}[P(x)]\}^2\big] \;=\; d \;.
\end{equation}
We know that $\tr_\mathrm{s}\!\big[\{\tr_\mathrm{a}[P]\}^2\big]\leq 1/d$, however, with equality only if 
$P=\ketbra{U}$ [via Eq.~(\ref{eq:ketU})], a maximally entangled state. Thus, since the normalization 
$\int_\mathscr{X} \d\tau(x)=D=d^2$ must be adhered to, Eq.~(\ref{eq:tightUC1}) can be satisfied only if 
$P(x)=\ketbra{U(x)}$, $\tau$-almost everywhere, for some function $U:\mathscr{X}\rightarrow\U(d)$.

We have established that all tight rank-one POVMs corresponding to $\UC$ have POVDs in the form $P(x)=\ketbra{U(x)}$ where 
$U:\mathscr{X}\rightarrow\U(d)$. It is thus natural to take $\mathscr{X}\subseteq\PU(d)$ and let
$U(x)$ denote a representative from the equivalence class of unitaries $x\in\PU(d)$ (as in Sec.~\ref{sec:designs}).
We will henceforth assume that this is the case. Now note that $\Tr(\Fs^2)=2$ under Eq.~(\ref{eq:ucsuperop}). This means
\begin{equation}
\frac{1}{d^4}\iint_\mathscr{X} \d\tau(x)\d\tau(y)\,|\tr[U(x)^\dag U(y)]|^{4} \;=\; 2 \;,
\end{equation}
given that $|\BraKet{P(x)}{P(y)}|^2=|\braket{U(x)}{U(y)}|^4=|\tr[U(x)^\dag U(y)]|^4/d^4$. In particular, if 
$\mathscr{X}$ is a finite set, then by Theorem~\ref{thm:welch}, $\mathscr{X}$ must be a weighted unitary 2-design with weight 
function $w(x)=\tau(x)/d^2$. In fact, Theorem~\ref{thm:welch} could easily be extended to any subset 
$\mathscr{X}\subseteq\PU(d)$ with the condition for equality in Eq.~(\ref{eq:welchbound}) (when $t=2$) replaced
by Eq.~(\ref{eq:tightUC}) in the following proposition (see e.g. Ref.~\cite[Theorem~6]{Scott06}).

\begin{prp}\label{prp:tightUC}
Let\/ $F:\mathfrak{B}(\mathscr{X})\rightarrow\H(\C^d\otimes\C^d)$ be a POVM with\/ $\supp(\Fs)=\Span(\UC)$ and assume
$\mathscr{X}\subseteq\PU(d)$. Then\/ $F$ is a tight rank-one POVM if and only if\/ $P(x)=\ketbra{U(x)}$ with the outcome
distribution\/ $(\mathscr{X},\tau/d^2)$ satisfying
\begin{equation}\label{eq:tightUC}
\frac{1}{d^2}\int_{\mathscr{X}}\d\tau(x)\, U(x)^{\otimes 2}\otimes \big(U(x)^{\otimes 2}\big)^\dag \;=\; \int_{\PU(d)}\!\!\d\muu(x)\, U(x)^{\otimes 2}\otimes \big(U(x)^{\otimes 2}\big)^\dag \;.
\end{equation}
That is, if\/ $\mathscr{X}$ is finite, then\/ $(\mathscr{X},\tau/d^2)$ is a weighted unitary 2-design.
\end{prp}

Weighted unitary 2-designs thus define the class of optimal measurements on the output state for linear ancilla-assisted process tomography of 
unital quantum channels. Proposition~\ref{prp:tightUC} and Corollary~\ref{cor:process} summarize this main result of the article.
By Corollary~\ref{cor:exist}, these measurements exist in all dimensions. One particularly interesting type is that specified by 
a complete set of MUUBs. This choice allows us to perform optimal ancilla-assisted process tomography through a series of orthogonal 
measurements on the output state. In all cases, the optimal reconstruction formula 
for the output state is [Eq.~(\ref{eq:reconstructtight}) with $D=d^2$ and $\delta=(d^2-1)^2+1$] 
\begin{equation}
\rho \;=\; (d^2-1)\int_\mathscr{X}\d p(x)\,\ketbra{U(x)} \;-\; \Big(1-\frac{2}{d^2}\Big)\,I\;,
\end{equation}
where $p(\mathscr{E})=\tr[F(\mathscr{E})\rho]=\int_\mathscr{E}\d\tau(x)\bra{U(x)}\rho\ket{U(x)}$ are the measurement outcome 
statistics. The corresponding unital channel follows from the Jamio{\l}kowski isomorphism [Eq.~(\ref{eq:Jam})]:
\begin{equation}
\mathcal{E} \;=\; (d^2-1)\int_\mathscr{X}\d p(x)\, U(x) \odot U(x)^\dag \;-\; \Big(d-\frac{2}{d}\Big)\,\Is\;.
\end{equation}

Finally, consider the case $\Qs=\GC$. Our condition for a tight rank-one POVM 
[Eq.~(\ref{tightdfneq3})] with $D=d^2$, $\delta=d^2(d^2-1)+1$ and $\Ps_{\Fs}=\Ps_{\GC}$ now becomes 
\begin{equation}\label{eq:gcsuperop}
\Fs \;=\;\int_\mathscr{X} \d\tau(x)\,\KetBra{P(x)} \;=\; \KetBra{\lambda_0\otimes\lambda_0}+\frac{1}{d^2}\mathop{\sum_{j>0}}_k\KetBra{\lambda_j\otimes\lambda_k}\;.
\end{equation}
But following the exact same procedure as in the unital case we find that the types of tight rank-one POVMs 
corresponding to $\GC$ must also satisfy Eq.~(\ref{eq:tightUC1}), and thus, we again have $P(x)=\ketbra{U(x)}$, 
$\tau$-almost everywhere, for some function $U:\mathscr{X}\rightarrow\U(d)$. In this case $\Tr(\Fs^2)=2-1/d^2$ under 
Eq.~(\ref{eq:gcsuperop}), however, which would violate Theorem~\ref{thm:welch}. Our only conclusion can be that 
tight rank-one POVMs with $\supp(\Fs)=\Span(\GC)$ do not exist. The lower bound on the error rate
[Eq.~(\ref{eq:errorrates})] still applies, but it is unattainable.

\section{Conclusion}
\label{sec:conclude}

In this article we have shown that weighted unitary 2-designs specify optimal measurements on the system-ancilla 
output state for ancilla-assisted process tomography of unital quantum channels (Corollary~\ref{cor:process} and 
Proposition~\ref{prp:tightUC}). Although existence is known in all dimensions 
(Corollary~\ref{cor:exist}), it remains to construct specific examples of these designs with sizes 
as close as possible to the lower bound (Theorem~\ref{thm:2design}). Complete sets of MUUBs are known in 
dimensions $d=2,3,5,7,11$, and form unweighted unitary 2-designs with sizes 
close to optimality. Each of these in fact specifies a minimal series of optimal orthogonal measurements (Theorem~\ref{thm:MUUBs}). 
Weighted unitary 2-designs of smaller size exist in dimension 2, however (see Table~\ref{tbl:dim2tdes}), and thus further 
reductions should be expected in higher dimensions. The optimization of the measurements used 
for ancilla-assisted process tomography of general quantum channels remains an open problem. 

\begin{acknowledgments} 
The author would like to thank Aidan Roy for helpful discussions. This work is supported by ARC and the State of Queensland.
\end{acknowledgments}

\appendix*
\section{Quantum operations}

Before describing quantum operations let us take a moment to set notation.
Following Caves~\cite{Caves99} (see also Refs.~\cite{Rungta00,Rungta01}) we write a linear operator $A$ in vector notation as $\Ket{A}$. The vector space of all such operators, 
$\End(\C^d)\cong\C^{d^2}$, equipped with the Hilbert-Schmidt inner product $\BraKet{A}{B}\coloneq\tr(A^\dag B)$, 
is a Hilbert space, where we think of $\Bra{A}$ as an operator ``bra'' and $\Ket{B}$ as an operator ``ket.''
Addition and scalar multiplication of operator kets then follows that for operators, e.g. $a\Ket{A}+b\Ket{B}=\Ket{aA+bB}$.
The usefulness of this notation becomes apparent when we consider linear maps on operators, i.e. superoperators. 
Given an orthonormal operator basis $\{E_k\}_{k=1}^{d^2}\subset\End(\C^d)$, $\BraKet{E_j}{E_k}=\delta_{jk}$, 
a superoperator $\mathcal{S}\in\End(\End(\C^d))\cong\C^{d^4}$ may be written in two different ways:
\begin{equation}\label{eq:super}
\mathcal{S}\;=\;\sum_{j,k}s_{jk}\,E_j\odot{E_k}^\dag\;=\;\sum_{j,k}s_{jk}\,\Ket{E_j}\Bra{E_k} \qquad\quad (s\in\C^{d^2 \times d^2})\;.
\end{equation}
The first representation illustrates the {\em ordinary\/} action of the superoperator, 
\begin{equation}
\mathcal{S}(A)\;\coloneq\;\sum_{j,k}s_{jk}E_j A{E_k}^\dag\;,
\end{equation} 
which amounts to inserting $A$ into the location of the `$\odot$' symbol. The second reflects the {\em left-right\/} 
action, 
\begin{equation}
\mathcal{S}\Ket{A}\;\coloneq\;\sum_{j,k}s_{jk}\Ket{E_j}\BraKet{E_k}{A}\;=\;\sum_{j,k}s_{jk}{E_j}\tr\big({E_k}^\dag{A}\big)\;,
\end{equation} 
where the superoperator acts on operators just like an operator on vectors. The identity superoperators relative to the 
ordinary and left-right actions are, respectively, $\mathcal{I}\coloneq I\odot I$ and $\Is\coloneq\sum_k\Ket{E_k}\Bra{E_k}$.
We also define $\Tr(\mathcal{S})\coloneq\sum_k\Bra{E_k}\mathcal{S}\Ket{E_k}$ and  
$\|\mathcal{S}\|\coloneq\sqrt{\Tr(\mathcal{S}^\dag\mathcal{S})}$. The latter is the Frobenius norm of $\mathcal{S}$ 
induced by its left-right action. Here $\mathcal{S}^\dag$ is the left-right adjoint, 
i.e. $\Bra{A}\mathcal{S}^\dag\Ket{B}\coloneq\Bra{B}\mathcal{S}\Ket{A}^*$, and $\mathcal{R}\mathcal{S}$ denotes 
the left-right composition of two superoperators: $(\mathcal{R}\mathcal{S})\Ket{A}\coloneq\mathcal{R}\Ket{B}$ where $\Ket{B}=\mathcal{S}\Ket{A}$.
Consult Refs.~\cite{Caves99,Rungta00,Rungta01} for analogous concepts relative to the ordinary action.

The particular choice of operator basis $\{E_k=E_{k_1k_2}\coloneq\ket{k_1}\bra{k_2}\}_{k_1,k_2=1}^{d}$,
where $\{\ket{k}\}_{k=1}^{d}$ is a fixed ``standard'' basis for $\C^d$, defines the so-called 
{\em Jamio{\l}kowski isomorphism\/}~\cite{Jamiolkowski72}. The matrix $s$ in Eq.~(\ref{eq:super}), 
now called the {\em process matrix\/}, then satisfies 
\begin{equation}
s_{jk} \;=\; s_{j_1j_2,k_1k_2} \;=\; d\,(\bra{j_1}\otimes\bra{j_2})\big[(\mathcal{S}\otimes\mathcal{I})(\ketbra{I})\big](\ket{k_1}\otimes\ket{k_2}) \;,
\end{equation}
where $\ket{I}\coloneq\sum_{k=1}^d \ket{k}\otimes\ket{k}/\sqrt{d}$. Note that $\Tr(\mathcal{S}^\dag\mathcal{S})=\tr(s^\dag s)$ 
in general, and thus, the Hilbert-Schmidt superoperator distance between $\mathcal{S}$ and $\mathcal{R}$ can be rewritten as 
$\|\mathcal{S}-\mathcal{R}\|=\|s-r\|$, where $r$ and $\mathcal{R}$ are related through 
Eq.~(\ref{eq:super}). The upshot of the current choice of operator basis is that when $\mathcal{S}$ and $\mathcal{R}$ 
are quantum channels, as in this article, then $s$ and $r$ specify standard-basis matrix elements of output quantum states 
with fixed input $\ketbra{I}$.

A superoperator $\mathcal{S}$ is called {\em positive\/} if it maps positive operators to positive operators under its ordinary action.
If, in addition, for any auxiliary system $\C^{d_\mathrm{a}}$ we have $(\mathcal{S}\otimes\mathcal{I})(A)\geq 0$ whenever 
$A\geq 0$, $A\in\End(\C^d\otimes\C^{d_\mathrm{a}})$, then $\mathcal{S}$ is called {\em completely positive\/}. 
Alternatively, $\mathcal{S}$ is completely positive if and only if $\Bra{A}\mathcal{S}\Ket{A}\geq 0$ for all 
$A\in\End(\C^d)$~\cite{Choi75}. That is, $\mathcal{S}$ is completely positive if and only if $\mathcal{S}^\dag=\mathcal{S}$ and 
$\mathcal{S}$ has nonnegative left-right eigenvalues. Diagonalizing, we see that $\mathcal{S}$ is completely positive if and only if it can be 
rewritten in an operator-sum form, called the {\em Kraus representation\/}~\cite{Kraus71,Kraus83,Choi75}:
\begin{equation}
\mathcal{S} \;=\; \sum_k B_k \odot {B_k}^{\dag} \;=\;\sum_k \KetBra{B_k}  \;,
\end{equation}
where the operators $B_k\in\End(\C^d)$ are called {\em Kraus operators\/}. A superoperator $\mathcal{S}$ is said to be 
{\em trace nonincreasing\/} if $\tr[\mathcal{S}(A)]\leq\tr(A)$ for all $A$, and moreover, {\em trace preserving\/} if 
$\tr[\mathcal{S}(A)]=\tr(A)$ for all $A$. Thus the Kraus operators together satisfy $\sum_k {B_k}^{\dag}B_k\leq I$ when 
$\mathcal{S}$ is trace nonincreasing and $\sum_k {B_k}^{\dag}B_k=I$ when $\mathcal{S}$ is trace preserving.

A {\em quantum operation\/} is a superoperator-valued measure 
$\mathcal{E}[\,\cdot\,]:\mathfrak{B}(\mathscr{X})\rightarrow\End(\End(\C^d))$ on an outcome set $\mathscr{X}$, 
which satisfies (1)~$\mathcal{E}[\mathscr{S}]$ is completely positive and trace nonincreasing for all 
$\mathscr{S}\in\mathfrak{B}(\mathscr{X})$ with $\mathcal{E}[\emptyset]=0$, 
(2)~$\mathcal{E}[\bigcup_{k=1}^\infty\mathscr{S}_k]=\sum_{k=1}^\infty \mathcal{E}[\mathscr{S}_k]$ 
for any sequence of disjoint sets $\mathscr{S}_k\in\mathfrak{B}(\mathscr{X})$, and (3) $\mathcal{E}[\mathscr{X}]$ is 
trace preserving. In this article we always take $\mathfrak{B}(\mathscr{X})$ to be the Borel $\sigma$-algebra. 
Quantum operations can be {\em nonselective\/} (e.g. channels), in which case there is only one output 
$\rho'=\mathcal{E}(\rho)\coloneq\mathcal{E}[\mathscr{X}](\rho)$ for each input $\rho$, but are generally {\em selective} 
(e.g. measurements), in which case the output $\rho'=\mathcal{E}[\mathscr{S}](\rho)/p(\mathscr{S})$ occurs with probability 
$p(\mathscr{S})=\tr\!\big(\mathcal{E}[\mathscr{S}](\rho)\big)$. To make the connection to quantum measurements simply note that 
$F(\,\cdot\,)\coloneq\sum_k {A_k(\,\cdot\,)}^{\dag}A_k(\,\cdot\,)$ is the POVM describing the outcome statistics of the 
measuring instrument $\mathcal{E}[\,\cdot\,]=\sum_k A_k(\,\cdot\,) \odot {A_k(\,\cdot\,)}^{\dag}$.

\end{document}